\documentclass[aps,prb, twocolumn,nofootinbib,groupedaddress,superscriptaddress,longbibliography,notitlepage, 10pt]{revtex4-2}

\usepackage{amsmath,amssymb}
\usepackage[normalem]{ulem}
\usepackage{graphicx}
\usepackage{longtable,multirow}
\usepackage{mathtools}
\usepackage{dsfont}
\usepackage{amsfonts}
\usepackage{xcolor}
\usepackage{url}
\usepackage{braket}
\usepackage{array}
\usepackage{booktabs}

\usepackage[colorlinks=true,linkcolor=teal,citecolor=teal,urlcolor=teal,hypertexnames=false]{hyperref}
\newcommand{\nocontentsline}[3]{}
\newcommand{\tocless}[2]{\bgroup\let\addcontentsline=\nocontentsline#1{#2}\egroup}

\def\ba#1\ea{\begin{align}#1\end{align}}
\def\bg#1\eg{\begin{gather}#1\end{gather}}
\def\bpm{\begin{pmatrix}}
\def\epm{\end{pmatrix}}
\def\bbm{\begin{bmatrix}}
\def\ebm{\end{bmatrix}}

\newcommand{\bs}[1]{\boldsymbol{#1}}

\newcommand{\teal}[1]{{\color{teal} #1}}

\allowdisplaybreaks

\begin{document}
\title{Composite Quantum Geometry and Semiclassical Dynamics}

\author{Henry Davenport}
\affiliation{Blackett Laboratory, Imperial College London, London SW7 2AZ, United Kingdom}
\author{Yoonseok Hwang}
\affiliation{Blackett Laboratory, Imperial College London, London SW7 2AZ, United Kingdom}
\author{Johannes Knolle}
\affiliation{Technical University of Munich, TUM School of Natural Sciences, Physics Department, 85748 Garching, Germany}
\affiliation{Munich Center for Quantum Science and Technology (MCQST), Schellingstr. 4, 80799 M\"unchen, Germany}
\author{Frank Schindler}
\affiliation{Blackett Laboratory, Imperial College London, London SW7 2AZ, United Kingdom}

\begin{abstract}
We derive semiclassical equations of motion for general composite bound states in insulators and semiconductors, covering excitations such as excitons and trions. For neutral composites we find that a uniform external electric field does not couple to a Berry curvature term, contrary to the naive expectation from single-electron dynamics. Instead, a distinct quantum geometric quantity appears generically in the equations of motion. This quantity is the \emph{difference} between inequivalent Berry connections that can be defined for the composite, generalising the concept of the quantum geometric dipole previously studied for excitons. In the case of charged composites such as trions, we find an additional Berry curvature contribution to the equations of motion. As we demonstrate, however, there is an infinite family of inequivalent composite Berry curvatures, and so care must be taken to make the correct choice that describes the physical dynamics. We explain how this choice should be made dependent on the definition of a spatial centre for the composite. We end by discussing composite dynamics that have no single-electron counterpart. We find that trions in magic-angle twisted bilayer graphene undergo a transverse drift under an applied electric field and that this is driven not only by the Berry curvature contribution but also by the quantum geometric dipole. The interplay of these two geometric contributions further imprints itself on the trion's internal dynamics, causing its dipole moment to oscillate in time.
\end{abstract}

\maketitle

\let\oldaddcontentsline\addcontentsline
\renewcommand{\addcontentsline}[3]{}

\twocolumngrid
\teal{\it Introduction---} Topology is fundamental to our understanding of modern condensed matter physics, and is essential in explaining both quantized transport and robust edge modes~\cite{HasanKaneRMP10, QSHEffectGrapheneMele2005, QSHEffectBernevig2006, WenZooTopPhases}. One important application is band topology which classifies the phases of non-interacting electrons in the presence of both crystalline and local symmetries~\cite{
FuTCI11, 
topocrysInsulatorsRobertJan2013, 
Ryu_2010, 
Ashvin230SINatComm17, 
AndreiTQC17, 
hwang2025stablerealspaceinvariantstopology, RobertJanPRX17, Kitaev_2009, hwang2025stablerealspaceinvariantstopology}. Given this success, one area of increasing interest is to extend band topology to interacting systems~\cite{IntGroundstate1, IntGroundstate2, IntGroundstate3, IntGroundstate4}. Interactions can give rise to entirely new classes of excitations, which may themselves carry non-trivial topology. This possibility has attracted particular attention in moiré systems, which offer a highly tunable platform for probing the interplay between single-particle band topology and interaction-driven excitation topology~\cite{topExcitonsMoire, trionsSchindler, ExcitonsBiasedBilayerGraphene, RhombohedralGrapheneExcitons, ExcitonTopologyPaperKwan, ExcitonTopologyPaperUchoa, yang2026gianthelicalexcitondipole, ledwith2025exoticcarriersconcentratedtopology, TrionsEslam}.

Recent work on the topology of excitons, bound states of electrons and holes, has uncovered a wealth of new features that extend beyond non-interacting electron topology~\cite{ExcitonBerryology, ExcitonTopologyDavenport, RhombohedralGrapheneExcitons, ExcitontopologySlager, ExcitonTopologyPaperKwan, ExcitonTopologyPaperQian, ExcitonTopologyPaperUchoa, FuChernExcitons, QuantumGeometricExcitonDrift, hwang2026stablewavefunctionzerosindicate, thompson2025topologicallyenhancedexcitontransport, paiva2024shiftpolarizationexcitonsquantum, PhysRevB.96.161101, eto2026oddparitymagnonshaldanehubbardmodel}. For example, excitons can have non-trivial topology even if the constituent electrons are trivial~\cite{ExcitonTopologyDavenport}. Even the fundamental building block of topological invariants, the Berry connection, is non-trivial for excitons. Unlike electrons, which possess a unique Berry connection, excitons have infinitely many~\cite{QuantumGeometricExcitonDrift, ExcitonBerryology}. This richness has a natural origin: for electrons, a celebrated result links the centres of maximally localised Wannier functions (MLWFs) to the Berry phase~\cite{MLWFVanderbilt}. Since excitons are composite bound states of electron-hole pairs, one can localise either the electron or the hole constituent, or some linear combination of their positions. Each of these choices yields a different set of Wannier centres and a distinct Berry connection~\cite{ExcitonTopologyDavenport, NeatonExcitonMLWFsPRB23, ExcitonBerryology}. Despite this complication, any exciton Berry connection can be written as a linear combination of the electron-localising and hole-localising ones~\cite{ExcitonBerryology}.

In this work we show that these features are general properties of composite excitations on top of non-interacting ground states. We extend the modern theory of polarisation to arbitrary composite particles and show that there is an infinite family of Berry connections. These are linear combinations of as many Berry connections as there are \emph{distinguishable} particles in the composite excitation. The differences between these Berry connections are gauge invariant and extend the \emph{quantum-geometric dipole} (QGD)~\cite{QuantumGeometricExcitonDrift, ManyBodyQuantumGeometricDipole, chen2026quantumgeometricdipoletopologicalboost, mendez2026signaturesquantumgeometricdipole, ExcitonBerryology} to arbitrary composites. We show that these quantities modify the dynamics of the composite particles by calculating their semiclassical equations of motion (EOM). 

We also analyse the dynamics of composite particles and show how these go beyond the possible \emph{single} particle dynamics. We demonstrate that the quantum geometric exciton drift of Ref.~\cite{QuantumGeometricExcitonDrift} is not specific to excitons but is a general feature of composite particle motion. Turning to trions in magic-angle twisted bilayer graphene (MATBG) at neutral filling, we find that the trion quantum geometric dipole forms a helix in momentum space, and that this leads to the electron and hole Berry curvatures differing substantially. Together these features lead to rich dynamics under an applied electric field: the electrons move perpendicular to the applied field due to Berry curvature, whilst the hole drift primarily arises from the QGD contribution to the EOM. These two contributions also modify the internal dynamics of the trion and lead to its dipole oscillating in time.

\teal{\it Bound state Berryology---} We study generic bound states in insulators. We express bound state wave functions in terms of creation operators $c^\dagger_{\bs{k}, \alpha}$ for states at momentum $\bs{k}$ and species label $\alpha$, \emph{e.g.} these could label the different electronic bands such as valence and conduction bands, or the spin or valley indices. The non-interacting insulating ground state is formed from the subset of filled \emph{valence} bands (denoted by ``$\mathrm{val}$''), so that $\ket{\mathrm{GS}} = \prod_{\bs{k}, \alpha \in \mathrm{val}} c^\dagger_{\bs{k}, \alpha} \ket{0}$. Assuming translational symmetry, we have bound state wave functions $\ket{\bs{p}}$ as a function of total momentum $\bs{p}$; these are eigenstates of a many-body Hamiltonian $\hat{\mathcal{H}}_0$ with energy $E(\bs{p})$. The bound state can be made of different species of particles. We write an arbitrary bound state as a superposition of states with the envelope wave function $\phi^{(\bs{k})}$,
\begin{equation}
\ket{\bs{p}}= \mathcal{
N}\sum_{\bs{K} = \bs{p}} \phi^{(\bs{k})}c^\dagger_{\bs{k}_1, \alpha_1}c^\dagger_{\bs{k}_2, \alpha_2}\dots c_{\bs{k}_{N-1}, \alpha_{N-1}}c_{\bs{k}_{N}, \alpha_{N}}\ket{\mathrm{GS}},\label{eq:boundstateWavefunction}
\end{equation}
where $\mathcal{N}$ is a normalisation factor.
The momentum labels of all particles and hole operators are denoted using the shorthand notation $(\bs{k}) \equiv (\bs{k}_1, \bs{k}_2, \dots \bs{k}_N)$. We sum over states with total momentum $\bs{K} = \bs{p}$ where $\bs{K}$ is the sum of the momenta of the particles (represented by $c^\dagger_{\bs{k}, \alpha}$) and holes  (represented by $c_{\bs{k}, \alpha}$). Note that in this sum the hole operator $c_{\bs{k}, \alpha}$ carries momentum $-\bs{k}$ as it removes a particle with momentum $\bs{k}$.
This ansatz covers a broad range of excitations — from conventional ones such as single electrons, excitons, magnons, trions, and biexcitons (including cases with multiple particles of the same species), to  exotic excitations like phonons in anomalous Hall crystals~\cite{hirsbrunner2026topologicalphononsanomaloushall}.

As shown in Ref.~\cite{ExcitonBerryology}, there are an infinite number of exciton Berry connections, constructible as linear combinations of two physically distinct connections. We now show that this is generally true: composite particles have as many \emph{genuinely} distinct Berry connections as there are distinct species of particles in the composite state (note this is the number of distinct species, not the total number of particles and holes). We first recall the modern theory of polarisation for non-interacting electrons before generalising to many-body composite states. Here the eigenstates of the projected position operator are the MLWFs (in 1D) and the eigenvalues are the centres of the MLWFs~\cite{MLWFVanderbilt}. The shift of the MLWF centres from the unit cell centres is the electronic Berry phase (divided by $2\pi$). This idea can be generalised to bound states to extract Berry phases and connections. Firstly, for composite particles, we can define a position operator $\hat{\bs{R}}^\alpha$ for each distinct species $\alpha$ in the bound state. We construct this position operator by taking the full position operator and rewriting it in terms of the electronic Bloch functions $\psi^{\bs{k}, \beta}_{\bs{R}, i}\equiv e^{\mathrm{i}\bs{k}\cdot \bs{R}} u^{\bs{k}, \beta}_{i} /\sqrt{\mathcal{V}}$ (labelling the unit cells by $\bs{R}$, orbitals within the unit cell by $i$ and species/non-interacting bands by $\beta$, while the normalisation factor $\mathcal{V}$ equals the total number of unit cells). This gives,
\begin{align}
\hat{\bs{R}} &= \sum_{\bs{R}, i} \bs{R} c^\dagger_{\bs{R}, i} c_{\bs{R}, i}\\
&= \sum_{\bs{k}, \bs{k}', \beta, \beta', \bs{R}, i} \bs{R} \:\bar\psi^{\bs{k}, \beta}_{\bs{R}, i} \psi^{\bs{k}', \beta'}_{\bs{R}, i} c^\dagger_{\bs{k}, \beta} c_{\bs{k}', \beta'},
\end{align}
If species $\alpha$ is a particle (not hole) we construct $\hat{\bs{R}}^\alpha$ by dropping terms which depend on other species to give,
\begin{equation}
\hat{\bs{R}}^\alpha = \frac{1}{N_\alpha}\sum_{\bs{k}, \bs{k}', \bs{R}, i} \bs{R} \:\bar\psi^{\bs{k}, \alpha}_{\bs{R}, i} \psi^{\bs{k}', \alpha}_{\bs{R}, i} c^\dagger_{\bs{k}, \alpha} c_{\bs{k}', \alpha},\label{eq:particlePositionOperator}
\end{equation}
where we have divided by the number of particles of species $\alpha$ in the bound state ($N_\alpha$) to obtain an average position. 
For holes we instead define, 
\begin{equation}
\hat{\bs{R}}^\alpha = \frac{1}{N_\alpha}\sum_{\bs{k}, \bs{k}', \bs{R}, i} \bs{R} \:\psi^{\bs{k}, \alpha}_{\bs{R}, i} \bar \psi^{\bs{k}', \alpha}_{\bs{R}, i} c_{\bs{k}, \alpha} c^\dagger_{\bs{k}', \alpha}.\label{eq:holePositionOperator}
\end{equation}
For each species $\alpha$ in the bound state and in 1D, we define a projected position operator $\hat P \hat R^\alpha \hat P$ where $\hat P = \sum_{p} \ket{p}\!\bra{p}$ is the projector onto a particular bound state band. The eigenvectors of this operator are the Wannier states which maximally localise the average position of the $\alpha$-particles~\cite{MLWFVanderbilt, ExcitonBerryology}. Just as in the electronic case, the eigenvalues are given by a Berry phase (divided by $2\pi$). This is the average position of the $\alpha$-particles within the MLWFs. By a direct calculation of the eigenvalues of $\hat P \hat R^\alpha \hat P$ we extract a Berry connection (Appendix~\ref{sec:BoundStateBerry}). Generalising this Berry connection to arbitrary dimensions we obtain, 
\begin{equation}
\bs{\mathcal{A}}^\alpha (\bs{p}) \equiv \sum_{{\bs{K} = \bs{p}}}\left(\nu_\alpha\: \bar \phi^{(\bs{k})}\mathrm{i} \nabla_{\bs{k}_n}\phi^{(\bs{k})} + |\phi^{(\bs{k})}|^2 \bs{A}_{\mathrm{elec}, \alpha}(\bs{k}_n)\right).
\end{equation}
The $\bs{k}_n$-derivative is the momentum derivative for particle $n$, which should be a particle of species $\alpha$, in the bound state of Eq.~\eqref{eq:boundstateWavefunction}. (It does not matter which specific $n$ is chosen as long as it belongs to species $\alpha$.) This has a contribution from the electronic Berry connection for species $\alpha$ \emph{i.e.} $\bs{A}_{\mathrm{elec}, \alpha}(\bs{k}_n) = \bra{u^{ \bs{k}_n, \alpha} }\mathrm{i}\nabla_{\bs{k}_n}\ket{u^{ \bs{k}_n, \alpha}}$ where $\ket{u^{ \bs{k}_n, \alpha}}$ is the cell-periodic Bloch function. The $\nu_\alpha$ is $+1$ for a particle and $-1$ for a hole. (More practical gauge invariant Wilson loop formulations of the composite Berry phases are in Appendix ~\ref{sec:BoundStateBerry}). Applying this construction to each constituent species yields a distinct Berry connection, and any linear combination of these is itself a valid Berry connection. This construction naturally generates an infinite family of connections, mirroring the structure previously found for excitons~\cite{ExcitonBerryology}. These Berry connections are genuinely distinct (\emph{i.e.} by more than just a gauge transformation) so Berry phases calculated using these different Berry connections have different physical meanings. 
 
The non-uniqueness of the Berry connection has another remarkable consequence, in that there are a host of new gauge invariant quantities that can be formed. Each of these connections is a genuine connection so it transforms covariantly \emph{i.e.} when $\ket{\bs{p}}\rightarrow e^{\mathrm{i}\theta(\bs{p})}\ket{\bs{p}}$ the connections transform as $\bs{\mathcal{A}}^\alpha (\bs{p}) \rightarrow \bs{\mathcal{A}}^\alpha (\bs{p})  - \nabla_{\bs{p}}\theta(\bs{p})$. Since all connections belonging to different particle species $\alpha$ transform identically, their pairwise differences are gauge invariant, giving a set of gauge invariant quantities,
\begin{equation}
    \bs{\mathcal{D}}^{\alpha, \beta}(\bs{p}) = \bs{\mathcal{A}}^{\alpha}(\bs{p})-\bs{\mathcal{A}}^{\beta}(\bs{p}).
\end{equation}
For $N$ distinguishable sets of particles/holes in the bound state, we have $N-1$ linearly independent gauge invariant quantities of this form. (A more practical expression for $\bs{\mathcal{D}}^{\alpha, \beta}(\bs{p})$ that doesn't require a smooth gauge is given in Appendix ~\ref{sec:PhysicalMeaningofD}).

For generic composite particles in insulators, we see that there are many new gauge invariant quantities that arise from the composite nature of the excitations. The \emph{quantum-geometric dipole} (QGD) studied for two particle composites~\cite{QuantumGeometricExcitonDrift, ManyBodyQuantumGeometricDipole, chen2026quantumgeometricdipoletopologicalboost} therefore generalises naturally to arbitrary composites. When there is one particle of each species and the two species carry equal and opposite charges, $\bs{\mathcal{D}}^{\alpha,\beta}(\bs{p})$ multiplied by the electric charge reduces to the physical dipole moment. More generally, however, it represents the average separation between the mean positions of species $\alpha$
and $\beta$ (see Appendix~\ref{sec:PhysicalMeaningofD}), and remains a well-defined gauge-invariant quantity even when the two species carry charges of the same sign or appear in unequal numbers. With this understanding, we will refer to $\bs{\mathcal{D}}^{\alpha,\beta}(\bs{p})$
as the QGD throughout.

We now derive the equations of motion for a general composite and find that all of the QGDs enter directly.

\teal{\it General equations of motion for bound states---}
We calculate the EOM for bound states in insulators. We begin with the many-body Hamiltonian and add a constant electric field,
\begin{equation}
\hat{\mathcal{H}}' = \hat{\mathcal{H}}_0 - \sum_{\bs{R}, i} q \bs{E}\cdot \bs{R} c^\dagger_{\bs{R}, i}c_{\bs{R}, i}
\end{equation}
where $q$ is the electric charge (note that therefore we are restricting to bound states of particles all with the same elementary charge $q$, where $q = -|e|$ for electrons). It is easier to calculate the EOM in a different electromagnetic gauge. We make the gauge transformation $\hat{U}(t) =\exp(-\mathrm{i}q\bs{E}\cdot \hat{\bs{R}}t)$. This acts on the creation operator for momentum $\bs{k}$ and orbital $i$ as, $\hat{U}^\dagger(t) c^\dagger_{\bs{k}, i} \hat{U}(t) = c^\dagger_{\bs{k}+q\bs{E} t, i}$. Under such a transformation, the Hamiltonian transforms to the time \emph{dependent} $\hat{\mathcal{H}}(t)= \hat U(t) \hat{\mathcal{H}}' \hat U^\dagger(t) + \mathrm{i} \partial_t\left[\hat U(t)\right] \hat U^\dagger(t)$ (see Appendix.~\ref{sec:UnitaryTransformation}). The second term in this transformation cancels with the scalar potential so the time-dependent Hamiltonian is simply $\hat{\mathcal{H}}(t) = \hat U(t) \hat{\mathcal{H}}_0 \hat U^\dagger(t)$.

We calculate the time derivative of the position operator for each particle species [Eqs.~\eqref{eq:particlePositionOperator}~\&~\eqref{eq:holePositionOperator}] to obtain the EOM. Firstly, we calculate the time evolution of the states using the adiabatic theorem \emph{i.e.} we assume that the bound state band of interest is well separated from other bands. The bound state follows the instantaneous eigenstates of $\hat{\mathcal{H}}(t)$ up to a phase. From the form of $\hat{\mathcal{H}}(t)$, we can construct instantaneous eigenstates for a bound state with conserved initial (=canonical) momentum $\bs{p}_0$ that read $\ket{\psi_0(\bs{p}_0, t)} \equiv \hat{U}(t)\ket{\bs{p}(t)}$, where $\ket{\bs{p}}$ from Eq.~\eqref{eq:boundstateWavefunction} denotes an eigenstate of $\hat{\mathcal{H}}_0$. The total \emph{kinetic} momentum of the bound state is then $\bs{p}(t) = \bs{p}_0 + \sum_{\alpha} N_\alpha \nu_{\alpha} q\bs{E}t$ since each particle in the bound state has a rate of change of momentum $q\bs{E}$, and each hole has a rate of change of momentum $-q\bs{E}$~(Appendix \ref{sec:InstantaneousEigStates}).

The adiabatic approximation stipulates that the bound state wave function at time $t$ follows the instantaneous eigenstate $\ket{\psi_0(\bs{p}_0, t)}$ but with an accumulated phase, $\ket{\psi(\bs{p}_0, t)} = \gamma(\bs{p}_0, t) \ket{\psi_0(\bs{p}_0, t)}$, where
\begin{equation}
\gamma(\bs{p}_0, t) = e^{-\mathrm{i}\int_0^t dt'\left[E(\bs{p}(t')) -\mathrm{i}\braket{\psi_0(\bs{p}_0, t')|\dot \psi_0(\bs{p}_0, t')}\right]}.
\end{equation}
We now calculate the time derivative of the expectation value of the position operator for each species in the time-evolved state $\ket{\psi(\bs{p}_0, t)}$, noting that the gauge transformation requires us to use the gauge transformed position operators $\hat{\bs{R}}^\alpha \rightarrow \hat{U}(t)\hat{\bs{R}}^\alpha \hat{U}^\dagger(t)$. The $\hat{U}(t)$ in $\ket{\psi(\bs{p}_0, t)}$ and in the gauge transformed position operator however cancel to give
\begin{equation}
\frac{\mathrm{d}\langle \hat{\bs{R}}^\alpha\rangle}{\mathrm{d}t}=\frac{\mathrm{d}}{\mathrm{d}t}\left[\bar \gamma \bra{\bs{p}(t)} \hat{\bs{R}}^\alpha\gamma \ket{\bs{p}(t)}\right],
\end{equation}
where $\bar \gamma \equiv (\gamma)^*$ and we drop the $\bs{p}_0, t$ dependence to avoid notational clutter.

In the thermodynamic limit, the position operator for species $\alpha$ acts like the derivative of the momentum of the $\alpha$ particle species (Appendix~\ref{sec:PosOperatorDerivative}) meaning that the Berry connection for species $\alpha$ enters the EOM. By the product rule we find that,
\begin{align}
\frac{\mathrm{d}\langle \hat{\bs{R}}^\alpha\rangle}{\mathrm{d}t}&=\mathrm{i}\frac{\mathrm{d}}{\mathrm{d}t}\left(\bar\gamma \nabla_{\bs{p}_0}\gamma\right) + \frac{\mathrm{d}}{\mathrm{d}t}\bs{\mathcal{A}}^\alpha(\bs{p}).
\end{align}
Plugging in the expression for $\gamma$ we obtain,
\begin{align}
\frac{\mathrm{d}\langle \hat{\bs{R}}^\alpha\rangle}{\mathrm{d}t}&=  \nabla_{\bs{p}}E - \dot{\bs{p}}\times \bs{\Omega}^{\alpha}(\bs{p})\label{eq:EOMStep3}\\*&\quad -  \nabla_{\bs{p}}\left(\mathrm{i} \braket{\psi_0| \partial_t\,\psi_0} - \dot{\bs{p}}\cdot \bs{\mathcal{A}}^\alpha(\bs{p})\right).\nonumber
\end{align}
We have defined the Berry curvature $\bs{\Omega}^{\alpha}(\bs{p})\equiv \nabla_{\bs{p}}\times \bs{\mathcal{A}}^\alpha(\bs{p})$. The EOM for electrons includes only the first line of Eq.~\eqref{eq:EOMStep3} but composite particles generically have a second set of electric field dependent terms~\cite{BerryPhaseElectrons}. 

We can simplify these additional terms using $\mathrm{i}\nabla_{\bs{p}}\braket{\psi_0| \partial_t\,\psi_0} = \nabla_{\bs{p}}[\sum_\beta N_\beta \nu_\beta q\bs{E}\cdot \bs{\mathcal{A}}^\beta(\bs{p})]$ (proof in Appendix~\ref{sec:groundstatedependencedropsout}). This follows from the chain rule: the rate of change of momentum of a particle of species $\beta$ is $\nu_\beta q\bs{E}$ and $\bs{\mathcal{A}}^\beta(\bs{p})$ involves a derivative with respect to the same momentum. (In Appendix~\ref{sec:groundstatedependencedropsout} we show that the ground state contributions to this time derivative vanish for the special case of non-interacting ground states.) We see that the QGDs naturally appear in the EOM since the time derivative can be expressed as a sum over \emph{all} the different species' Berry connections, whereas the position operator introduces just $\bs{\mathcal{A}}^{\alpha}(\bs{p})$. Recalling that $\dot{\bs{p}} = \sum_{\alpha} N_\alpha \nu_\alpha q \bs{E}$, we  arrive at the EOM for an arbitrary bound state,
\begin{equation}
\frac{\mathrm{d}\langle \hat{\bs{R}}^\alpha\rangle}{\mathrm{d}t}  =  \nabla_{\bs{p}}E - \dot{\bs{p}}\times \bs{\Omega}^{\alpha}(\bs{p}) +  \nabla_{\bs{p}}\bigg[q \bs{E}\cdot\sum_\beta N_\beta\nu_\beta \bs{\mathcal{D}}^{\alpha, \beta}(\bs{p})\bigg],\label{eq:FinalEOM}
\end{equation}
which has contributions from the bound state dispersion, Berry curvatures and QGDs.
In this derivation, there are only two approximations, firstly we have used the adiabatic approximation and secondly we have restricted the EOM to one momentum. This latter approximation can be easily dropped by considering a wave packet which is a superposition of different $\ket{\bs{p}}$ with weighting factors $a_{\bs{p}}$, the resulting EOM are simply Eq.~\eqref{eq:FinalEOM} weighted by $|a_{\bs{p}}|^2$ (see Appendix \ref{sec:PosOperatorDerivative}).

\teal{\it Examples---} We now discuss some  examples of the EOMs and the interesting dynamics that are possible. We begin with a single electron quasiparticle excitation above the ground state. Since the excitation contains only one species of particle, there is only one Berry connection and no pairs of connections to take the difference between, so the QGD contribution vanishes from Eq.~\eqref{eq:FinalEOM}. We therefore recover the celebrated electronic semiclassical EOM with anomalous Berry curvature term~\cite{BerryPhaseElectrons}. (In Appendix~\ref{sec:ElectronEOM} we provide a further simple derivation of the single electron EOM).

There are two particularly interesting features of the general EOM which we now explore. First of all we consider the QGD term which only appears for composites with different species constituents. The QGD terms in the general EOM take the form $\nabla_{\bs{p}}\left[q \bs{E}\cdot \bs{\mathcal{D}}^{\alpha, \beta}(\bs{p})\right] $. This can produce a number of different interesting dynamics beyond that contributed by the Berry curvature term. Notably, whereas the Berry curvature term contributes only a velocity perpendicular to the applied field, this quantity can also give a velocity parallel to it. For neutral composites, $\dot{\bs{p}} = 0$ and so, beyond the standard dispersion contribution $\nabla_{\bs{p}}E$, only the QGD contributes to the equations of motion. Although the Berry curvature term is absent, transverse drift analogous to that produced by Berry curvature can still arise due to the QGD. For example this occurs when the QGD has a helical texture in the Brillouin zone (BZ) \emph{i.e.} $\bs{\mathcal{D}}^{\alpha, \beta}(\bs{p}) \sim \bs{p}\times \hat{\bs{z}}$ in 2D with $\hat{\bs{z}}$ out of the plane. In this case, when we apply an electric field, the QGD term in the EOM is exactly perpendicular to $\bs{E}$, $\nabla_{\bs{p}}\left[q \bs{E}\cdot \bs{\mathcal{D}}^{\alpha, \beta}(\bs{p})\right] \sim \bs{E}_{\perp}$
where $\bs{E}_{\perp}\cdot \bs{E} = 0$. This drift is independent of the species' Berry curvature and so an anomalous perpendicular velocity can occur even when the Berry curvature vanishes. Our general EOM therefore predicts that the QGD induced drift predicted for excitons (and termed the quantum geometric exciton drift in Ref.~\cite{QuantumGeometricExcitonDrift}) extends to general composites. One interesting consequence of this is that, in the presence of a sufficiently smooth confining potential, composite particles can circulate perpendicular to the edge. This is much like the chiral edge mode associated with a non-zero Chern number but can occur when the Chern number and even Berry curvature vanishes. Finally, we note that the particular QGD helix texture we described above does indeed occur in real materials. In Ref.~\cite{yang2026gianthelicalexcitondipole} the excitons in 2.1$^\circ$ twisted $\mathrm{MoTe}_2$ at a hole filling factor of one are shown to have exactly this helical QGD texture and so they should exhibit a perpendicular drift upon applied electric field. 

\begin{figure}[t]
\centering
\includegraphics[width=0.5\textwidth]{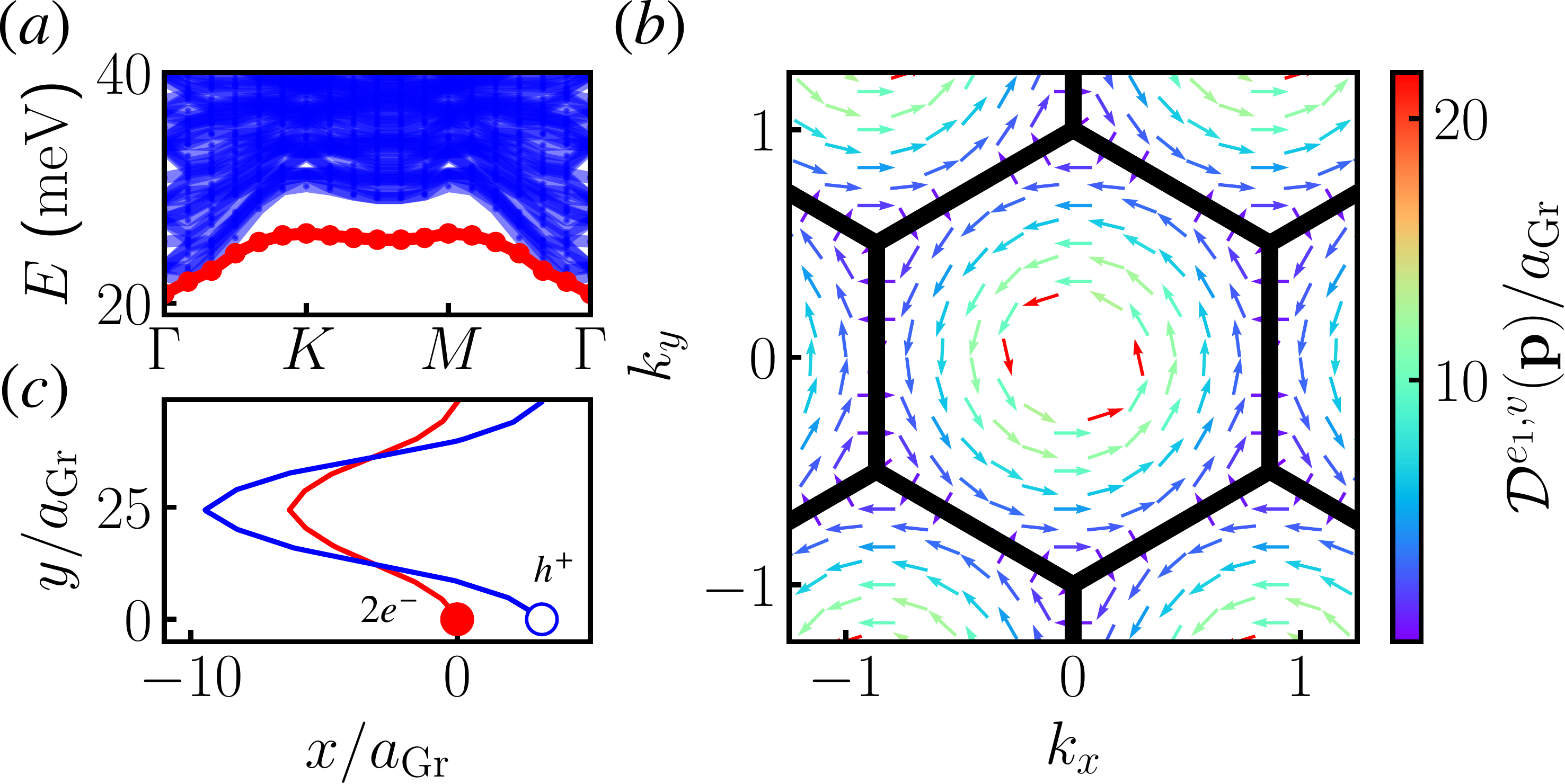}
\caption{Trions in MATBG at filling $\nu = 0$, in the chiral limit $w_0/w_1 = 0$ with gate distance $\xi = 20\:\mathrm{nm}$~\cite{trionsSchindler}. $(a)$ Trion dispersion with gapped trion away from the $\Gamma$ point. $(b)$ QGD [$\bs{\mathcal{D}}^{e_1, h}(\bs{p})$] for the lowest energy bound trion. $(c)$ Dynamics of trion starting at $\boldsymbol{k} = (0, 0.75)$ with applied field in $\hat{\bs{x}}$ direction.}
\label{fig:TrionsTBG}
\end{figure}

A second notable feature of the EOM is that distinct Berry curvatures enter for each constituent particle. However, as discussed above, these Berry curvature terms vanish for charge-neutral composites in a uniform electric field. To isolate their effect, we therefore turn to the equations of motion for charged three-body composites — specifically trions, which consist of either two electrons and one hole, or two holes and one electron. We calculate the dynamics for trions in MATBG. This is a natural setting to study, since the trions in MATBG have been proposed as a route to understanding the material's enigmatic superconductivity~\cite{TrionsEslam, trionsSchindler, ledwith2025exoticcarriersconcentratedtopology}, and our results offer new insight into their field-driven dynamics. We focus on the two electrons (labelled $e_1/e_2$) and one hole ($h$) trion in the chiral limit of MATBG~\cite{trionsSchindler}. In this limit the trion is gapless at the $\Gamma$ point but well gapped at the moire BZ boundary [Fig.~\ref{fig:TrionsTBG}(a)]. Interestingly, we find that the hole Berry curvature $\bs{\Omega}^h$ is much smaller than the electron Berry curvatures $\bs{\Omega}^{e_1/e_2}$ (see Appendix~\ref{sec:TrionsMATBG}). The two QGDs $\bs{\mathcal{D}}^{e_1/e_2, h}(\bs{p})$ [see Fig.~\ref{fig:TrionsTBG}(b)] are equal in the chiral limit and form a helix in the BZ. Towards the $\Gamma$
point, the QGD magnitude grows, consistent with the trion becoming unbound and merging with the three-body continuum. The large differences in the hole and electron Berry curvatures are explained by the QGD helix: the difference of the electron and hole Berry curvatures is equal to the curl of the QGD which — for this helical texture — is large. Using these quantities we calculate the dynamics for a trion starting at $(k_x, k_y) = (0.75, 0)$ with an electric field applied in the $\hat x$ direction [Fig.~\ref{fig:TrionsTBG}(c)]. Naively, the small Berry curvature of the hole vs the electrons might be expected to unbind the trion: the electrons acquire a much larger transverse velocity from the Berry curvature compared to the hole, seemingly driving the bound state apart. However, the QGD terms precisely compensate for this imbalance — contributing a transverse velocity of their own, as we showed was possible above — and the composite remains bound. This is a manifestation of a general property of the EOM in Eq.~\eqref{eq:FinalEOM}: the QGD and the Berry curvatures enter the EOM in such a way that the composite always remains bound (Appendix~\ref{sec:BoundStatesStayBound}). The helical quantum geometric dipole texture further causes the trion's dipole moment to oscillate as the trion drifts in the $\hat y$ direction. The system thus exhibits an AC response despite the applied DC field — a distinctly many-body effect with no single-particle analogue — and the resulting oscillating dipole will emit radiation perpendicular to the plane of the MATBG, offering a clear experimental signature.

\teal{\it Discussion---} We have shown that composite excitations on top of non-interacting ground states have as many genuinely distinct Berry connections as there are distinguishable constituent particles/holes, and that their pairwise differences are gauge invariant, generalising the quantum geometric dipole previously studied for two-body composites. We find that all of these gauge-invariant quantities appear directly in the semiclassical EOM, connecting the underlying quantum geometry to physical observables. Beyond this, we have provided gauge-invariant expressions for topological invariants in terms of Wilson loops, a physical interpretation of composite particle topology, and a clear understanding of the generalised quantum geometric dipole — together opening up a systematic framework for studying the quantum geometry of arbitrary composite excitations.

One extension of our formalism is to non-uniform electric fields. In electronic systems, the quantum metric also contributes if the electric field is non-uniform~\cite{SemiclassicsQuantumMetric, SemiclassicsQuantumMetric2}. Since the quantum metric is related to the variance of the MLWFs, it is likely that there is an infinite family of quantum metrics for composite particles that enter the EOM.

We note that our bound state EOM extend to a number of additional interesting applications. One natural extension is to composites without definite particle number. Superpositions of trions and electrons have recently been predicted in pentalayer rhombohedral graphene~\cite{ledwith2025exoticcarriersconcentratedtopology}, where excitations interpolate between the two kinds of excitation with varying weights across the Brillouin zone. Since the position operator conserves particle number, our equations of motion apply to each component separately and can be combined as a weighted sum. The trion component experiences quantum geometric dipole corrections while the single-electron component does not, so the dynamics could vary significantly across the Brillouin zone. A second interesting application is the Goldstone modes of anomalous Hall crystals, which have recently been shown to harbour non-trivial topology~\cite{hirsbrunner2026topologicalphononsanomaloushall} using methods adopted from Ref.~\cite{ExcitonBerryology}. Studying their field-driven dynamics is particularly timely, as the response could serve as direct experimental confirmation of the predicted anomalous Hall crystal phase in rhombohedral graphene~\cite{hirsbrunner2026topologicalphononsanomaloushall, PhysRevLett.133.206503, PhysRevB.110.205130, AHCPaperNature1, PhysRevB.110.245115, PhysRevB.110.205124, PhysRevLett.133.206504, AHCNature2}, and our equations of motion, Eq.~\eqref{eq:FinalEOM}, provide a natural framework for doing so.


\teal{\it Note added---} During the preparation of this manuscript, the preprint Ref.~\cite{hu2026generalizedshiftvectorintrinsic} was released. The authors derive the QGD of a full
$N$-body state and connect it to physical observables such as electron-phonon mediated processes. Complementary to this, we resolve the QGD into contributions from the pairs of distinct species comprising the state, linking these to the infinite family of composite state Berry connections, and show that even these species-resolved QGDs enter directly into the equations of motion.

\teal{\it Acknowledgments---}
HD acknowledges support from the Engineering and Physical Sciences Research Council (grant number EP/W524323/1). This work was supported by a UKRI Future Leaders Fellowship MR/Y017331/1. JK acknowledges support from the Deutsche Forschungsgemeinschaft (DFG, German Research Foundation) under Germany’s Excellence Strategy–EXC– 2111–390814868, DFG grants No. KN1254/1-2, KN1254/2- 1, and TRR 360 - 492547816; as well as the Munich Quantum Valley, which is supported by the Bavarian state government with funds from the Hightech Agenda Bayern Plus. J.K. further acknowledges support from the Keck foundation.
We acknowledge support from the Imperial-TUM flagship partnership.

\bibliography{refs.bib}

\let\addcontentsline\oldaddcontentsline
\clearpage

\onecolumngrid
\begin{center}
\textbf{\large Supplemental Material for \\
``Composite Quantum Geometry and Semiclassical Dynamics''}
\end{center}

\setcounter{section}{0}
\setcounter{figure}{0}
\setcounter{equation}{0}
\renewcommand{\thefigure}{S\arabic{figure}}
\renewcommand{\theequation}{S\arabic{equation}}
\renewcommand{\thesection}{S\arabic{section}}
\tableofcontents

\hfill \\
\newpage
\section{Electron semiclassical equations of motion}\label{sec:ElectronEOM}
In this section we calculate the electronic semiclassical EOM. Our bound state EOM calculation is a generalisation of this method. 

We begin with the Bloch Hamiltonian,
\begin{equation}
\hat H_0 = \sum_{\bs{k}, \alpha, \beta}   \left[{h}(\bs{k})\right]_{\alpha, \beta} \ket{\bs{k}, \alpha} \bra{\bs{k}, \beta},
\end{equation}
where the $\alpha, \beta$ sums over the sites within the unit cell. Defining $\ket{\bs{k}, \alpha} = \frac{1}{\sqrt{\mathcal{V}}}\sum_{\bs{R}} e^{\mathrm{i}\bs{k}\cdot \bs{R}} \ket{\bs{R}, \alpha}$ labelling unit cells by $\bs{R}$. We write the eigenvalues of $h(\boldsymbol{k})$ are $E(\boldsymbol{k})$ for a given band. 

We add a constant electric field $\bs{E}$. We want a force $q \bs{E}$ on the electrons (which have charge $q$). Since $\bs{E} = -\nabla_{\bs{R}}\: V(\bs{R})$, we add a potential $V(\bs{R}) = -\bs{E}\cdot \bs{R}$. We therefore consider the Hamiltonian,
\begin{equation}
\hat H' = \hat H_0 - \sum_{\bs{R}, \alpha} q\bs{E}\cdot \bs{R} \ket{\bs{R}, \alpha}\bra{\bs{R}, \alpha},\label{eq:scalarPotentialHam}
\end{equation}
where $q$ is the electric charge. This Hamiltonian ($\hat H'$) is hard to work with since it explicitly breaks translational invariance. We can however make a \emph{time-dependent} unitary transformation to get rid of the scalar potential. This can be done because the electric field is $\bs{E} = -\nabla \phi - \frac{\partial \bs{A}}{\partial t}$, so the electric field can incorporated via a scalar potential or a time dependent vector potential $\bs{A} = -\bs{E}t$. To change the gauge from the former to the latter, we can use the time dependent unitary transformation $\hat{U}(t) = \sum_{\bs{R}, \alpha} \exp\left(-\mathrm{i}q\bs{E}\cdot \bs{R}t\right) \ket{\bs{R}, \alpha}\bra{\bs{R}, \alpha}$. If we transform our basis using $\hat U(t)$ then we have to transform our Hamiltonian to $\hat H(t)= \hat U(t)\hat H' \hat U^\dagger(t) + \mathrm{i} \partial_t\left[\hat U(t)\right] \hat U^\dagger(t)$ (see Appendix.~\ref{sec:UnitaryTransformation}). Calculating the second term we find,
\begin{align}
\mathrm{i} \partial_t\left[\hat U(t)\right] \hat U^\dagger(t) &= \mathrm{i}\sum_{\bs{R}, \alpha, \bs{R}', \alpha'}\left(-\mathrm{i}q\bs{E}\cdot \bs{R} \right)\exp\left(-\mathrm{i}q\bs{E}\cdot \bs{R}t\right) \ket{\bs{R}, \alpha}\bra{\bs{R}, \alpha}\\&\nonumber\quad\quad\quad\quad\quad\cdot\ \exp\left(\mathrm{i}q\bs{E}\cdot \bs{R}'t\right) \ket{\bs{R}', \alpha'}\bra{\bs{R}', \alpha'}\\
&=\sum_{\bs{R}, \alpha}q\bs{E}\cdot \bs{R} \ket{\bs{R}, \alpha}\bra{\bs{R}, \alpha}.
\end{align}
This term therefore exactly cancels the scalar potential term in Eq.~\eqref{eq:scalarPotentialHam}. Therefore $\hat H(t) = \hat U(t)\hat H_0 \hat U^\dagger(t)$. We can rewrite the unitary in momentum space as $\hat U(t) = \sum_{\bs{k}, \alpha} \ket{\bs{k}, \alpha}\bra{\bs{k}+q\bs{E}t, \alpha}$. Therefore,
\begin{align}
\hat H(t) &=  \hat U(t)\hat H_0 \hat U^\dagger(t)\\
&=  \sum_{\bs{k}, \bs{p}, \bs{p}'}\sum_{\alpha, \beta, \gamma, \gamma'}   \left[{h}(\bs{k})\right]_{\alpha, \beta} \ket{\bs{p}, \gamma}\braket{\bs{p}+q\bs{E}t, \gamma|\bs{k}, \alpha} \braket{\bs{k}, \beta|\bs{p}'+q\bs{E}t, \gamma'}\bra{\bs{p}', \gamma'}
\\
&=  \sum_{\bs{k}, \alpha, \beta}   \left[{h}(\bs{k})\right]_{\alpha, \beta}\ket{\bs{k}-q\bs{E}t, \alpha} \bra{\bs{k}-q\bs{E}t, \beta}
\\
&=\sum_{\bs{k}, \alpha, \beta}   \left[\hat h(\bs{k}+ q \bs{E}t)\right]_{\alpha,\beta}\ket{\bs{k}, \alpha} \bra{\bs{k}, \beta}.
\end{align}
The change of summation variable in the last step  can only be done in the thermodynamic limit where $\bs{k}$ is continuous. The time dependent $\hat H(t)$ is diagonal in $\bs{k}$ for all $t$.

We write the components of the eigenvectors of $h[\bs{k}_0+q\bs{E}t]$ as $u^{\bs{k}_0+q\bs{E}t}_{\alpha}$ where $\alpha$ labels the orbital. The Hamiltonian $\hat H(t)$ has instantaneous eigenstates,
\begin{align}
\ket{\psi_0(\bs{k}_0, t)} &= \sum_{\alpha} u^{\bs{k}_0+q\bs{E}t}_{\alpha}\ket{\bs{k}_0, \alpha}\\
&=\ket{\bs{k}_0} \otimes \ket{u_{\bs{k}_0+q\bs{E}t}}.
\end{align}
In the last step we have changed notation using $\ket{\bs{k}_0, \alpha} = \ket{\bs{k}_0} \otimes \ket{\alpha}$ and the Bloch vector defined as $\ket{u_{\bs{k}}} \equiv \sum_\alpha u^{\bs{k}}_\alpha \ket{\alpha}$. (Throughout this derivation we only consider one band so we have dropped the band dependence of $\ket{u_{\bs{k}}}$.)

The state $\ket{\psi_0(\bs{k}_0, t)}$ is the instantaneous eigenstate at the conserved momentum $\bs{k}_0$. We now use the adiabatic approximation. This means that under time evolution the wave function follows the instantaneous eigenstate of $\hat H(t)$ (at fixed $\bs{k}_0$) but with some additional phase, \emph{i.e.} \begin{equation}
\ket{\psi(\bs{k}_0, t)} = \gamma(\bs{k}_0, t) \ket{\psi_0(\bs{k}_0, t)}
\end{equation}
with
\begin{equation}
\gamma(\bs{k}_0, t) =  \exp\left[{-\mathrm{i} \int_{0}^t E(\bs{k}_0+q \bs{E}t') \mathrm{d}t'} -\int_0^t \bra{u_{\bs{k}_0+q \bs{E}t'}}\partial_{t'}\ket{u_{\bs{k}_0+q \bs{E}t'}} \mathrm{d}t'\right].
\end{equation}
We calculate the time derivative of the expectation value of the electronic position operator [$\frac{d}{dt} \braket{\psi(t)| \hat{\bs{R}}| \psi(t)}$].
The position operator turns into the gradient operator with respect to canonical momentum ($\mathrm{i}\nabla_{\bs{k}_0}$) that acts on the cell periodic bit of the wave function.

To show the equivalence between the position operator and a momentum derivative on the Bloch cell periodic $\ket{u}$, we have to form a wave packet out of states at different momenta $\bs{k}$. This is because we need to use integration by parts to move the derivative to act on $\ket{u}$ not $\ket{\bs{k}}$. We have a wave packet with prefactors $a_{\bs{k}}$ for each Bloch state,
\begin{equation}
\ket{\psi(t)} = \left(\frac{L}{2\pi}\right)^d\int_{\mathrm{BZ}} {\mathrm{d}^d\bs{k}} \:a_{\bs{k}} \gamma_{\bs{k}} \ket{\bs{k}} \otimes \ket{u_{\bs{k} + q \bs{E}t}}.
\end{equation}
The position operator acts as a derivative on $\ket{\bs{k}}$ (\emph{i.e.} $\bs{R}\ket{\bs{k}} = -\mathrm{i} \nabla_{\bs{k}}\ket{\bs{k}}$),
\begin{align}
\hat{\bs{R}}\ket{\psi(t)} &= -\mathrm{i}\left(\frac{L}{2\pi}\right)^d\int_{\mathrm{BZ}} {\mathrm{d}^d\bs{k}} \: \nabla_{\bs{k}}\ket{\bs{k}} \otimes a_{\bs{k}} \gamma_{\bs{k}}\ket{u_{\bs{k} + q \bs{E}t}}\\
&= \mathrm{i}\left(\frac{L}{2\pi}\right)^d\int_{\mathrm{BZ}} {\mathrm{d}^d\bs{k}} \: \ket{\bs{k}} \otimes \nabla_{\bs{k}}\left(a_{\bs{k}} \gamma_{\bs{k}}\ket{u_{\bs{k} + q \bs{E}t}}\right)
\end{align}
using integration by parts. Therefore,
\begin{align}
\braket{\psi(t)| \hat{\bs{R}}| \psi(t)} &= \mathrm{i}\left(\frac{L}{2\pi}\right)^d\int_{\mathrm{BZ}} {\mathrm{d}^d\bs{k}} \:\bar a_{\bs{k}} \bar \gamma_{\bs{k}}\bra{u_{\bs{k} + q \bs{E}t}}\nabla_{\bs{k}}\left(a_{\bs{k}} \gamma_{\bs{k}}\ket{u_{\bs{k} + q \bs{E}t}}\right)\\
&= \mathrm{i}\left(\frac{L}{2\pi}\right)^d\int_{\mathrm{BZ}} {\mathrm{d}^d\bs{k}} \: |a_{\bs{k}} |^2\bar \gamma_{\bs{k}}\bra{u_{\bs{k} + q \bs{E}t}}\nabla_{\bs{k}}\left( \gamma_{\bs{k}}\ket{u_{\bs{k} + q \bs{E}t}}\right)\\&\quad\nonumber+ \mathrm{i}\left(\frac{L}{2\pi}\right)^d\int_{\mathrm{BZ}} {\mathrm{d}^d\bs{k}} \: \bar a_{\bs{k}}\nabla_{\bs{k}} a_{\bs{k}}
\end{align}
We look at the time derivative of  $\braket{\psi(t)| \hat{\bs{R}}| \psi(t)}$ meaning that the second term in this can be ignored. We now assume that the wave-packet is sufficiently well localised in momentum so that we only have to consider one momentum $\bs{k}$. Then we obtain the following, 
\begin{align}
\frac{d}{dt} \left[\braket{\psi(\bs{k}_0, t)| \hat{\bs{R}}| \psi(\bs{k}_0, t)}\right] &= \mathrm{i}\frac{d}{dt}\left[\bar\gamma \bra{u}\left(\gamma\ket{u}\right)'\right]\label{eq:positionIsderivative}\\
&= \mathrm{i}\frac{d}{dt}\left(\bar\gamma\gamma'\right)+ \mathrm{i}\braket{  \dot u
|u'}+\mathrm{i}\braket{ u|\dot u'}.\label{eq:productRuleElectron}
\end{align}
Note that we only have Bloch states at momentum $\bs{k}_0 + q \bs{E}t$ at time $t$ so we simplify our notation $\ket{u_{\bs{k}_0 + q \bs{E}t}}\rightarrow\ket{u}$ and $\gamma_{\bs{k}}(t)\rightarrow\gamma$. We also write the derivative $\nabla_{\bs{k}_0}$ as $'$.

Using the definition of $\gamma$ we see that,
\begin{equation}
 \mathrm{i}\frac{d}{dt}\left(\bar\gamma\gamma'\right) = E' - \mathrm{i} \left(\braket{ u| \dot u}\right)'
\end{equation}
The rate of change of the kinetic momentum is $\dot{\bs{k}} = q\bs{E}$ so we can write, 
\begin{align}
\frac{d}{dt} \braket{\psi(t)| \hat{\bs{R}}| \psi(t)} &= E'+ \mathrm{i}\left(\braket{\dot u|u'}-\braket{  u'| \dot u}\right)\\
&= \nabla_{\bs{k}}E+\mathrm{i}\dot{\bs{k}}\cdot(\nabla_{\bs{k}}\bra{ u})\nabla_{\bs{k}}\ket{u}
-\mathrm{i}(\nabla_{\bs{k}}\bra{  u})\dot{\bs{k}}\cdot\nabla_{\bs{k}} \ket{u},\\
&= \nabla_{\bs{k}}E+\mathrm{i}\dot{\bs{k}}\cdot\nabla_{\bs{k}}(\bra{ u}\nabla_{\bs{k}}\ket{u})
-\mathrm{i}\nabla_{\bs{k}}\left(\bra{  u}\dot{\bs{k}}\cdot\nabla_{\bs{k}} \ket{u}\right),\\
&= \nabla_{\bs{k}}E+\dot{\bs{k}}\cdot\nabla_{\bs{k}}(\bs{A}(\bs{k}))
-\nabla_{\bs{k}}\left(\dot{\bs{k}}\cdot \bs{A}(\bs{k})\right),\\
&= \nabla_{\bs{k}}E- \dot{\bs{k}} \times \bs{\Omega}(\bs{k}),
\end{align}
using a vector calculus identity to express the EOM in terms of the Berry curvature, $\bs{\Omega}(\bs{k}) = \nabla_{\bs{k}}\times \bs{A}(\bs{k})$ with Berry connection $\bs{A}(\bs{k})\equiv \bra{u}\mathrm{i}\nabla_{\bs{k}}\ket{u}$. In addition, we have replaced the derivatives with respect to canonical momentum $\nabla_{\bs{k}_0}$ with those with respect to kinetic momentum $\nabla_{\bs{k}}$ since these are equal when acting on $E$ and $\ket{u}$.  

Without assuming that one $\boldsymbol{k}$ dominates, we instead arrive at the weighted semiclassical EOM,
\begin{equation}
\frac{d}{dt} \braket{\psi(t)| \hat{\bs{R}}| \psi(t)}= \left(\frac{L}{2\pi}\right)^d\int_{\mathrm{BZ}}\mathrm{d}^d \bs{k}\: |a_{\bs{k}}|^2 \left[\nabla_{\bs{k}}E- \dot{\bs{k}} \times \bs{\Omega}(\bs{k})\right].
\end{equation}

Lastly, we briefly comment on the identification of $\boldsymbol{k}_0$ with the canonical momentum. In classical electrodynamics, the canonical momentum for a single particle of charge $q$ is $\bs{k}_{\mathrm{can}} \equiv \bs{k}_{\mathrm{kin}}(t)  + q \boldsymbol{A}(t)$ where $\bs{k}_{\mathrm{kin}}(t)$ is the kinetic momentum and $\boldsymbol{A}(t)$ is the vector potential. For the case of $\boldsymbol{A}(t) = -\boldsymbol{E}t$ we get $\bs{k}_{\mathrm{can}} \equiv \bs{k}_{\mathrm{kin}}(t)  - q \boldsymbol{E}t$. The charged particle's kinetic momentum is $\bs{k}_{\mathrm{kin}}(t)  = \bs{k}_0 + q\boldsymbol{E}t$. Therefore the canonical momentum $\bs{k}_{\mathrm{can}} = \bs{k}_0$. We therefore see that the canonical momentum is conserved (as also required by Noether's theorem).

\newpage
\section{Arbitrary bound state wave functions}\label{sec:Normalisation}
We now consider the wave function for an arbitrary bound state. We have bound state wave functions $\ket{\bs{p}}$ as a function of their total momentum $\bs{p}$, these are eigenstates of a translationally-symmetric many-body Hamiltonian $\hat{\mathcal{H}}_0$ with energy $E(\bs{p})$. The bound state can be made of different species of particles. We write an arbitrary bound state as,
\begin{equation}
\ket{\bs{p}}= \mathcal{
N}\sum_{\bs{K} = \bs{p}} \phi^{(\bs{k})}c^\dagger_{\bs{k}_1, \alpha_1}c^\dagger_{\bs{k}_2, \alpha_2}\dots c_{\bs{k}_{N-1}, \alpha_{N-1}}c_{\bs{k}_{N}, \alpha_{N}}\ket{\mathrm{GS}}\label{eq:AppboundstateWavefunction}
\end{equation}
where $(\bs{k})$ represents the momentum labels of all particles and holes in the bound state \emph{i.e.} $(\bs{k}) = (\bs{k}_1, \bs{k}_2, \dots \bs{k}_N)$. The $\alpha_i$ are the species labels \emph{e.g.} this could be band labels such as valence or conduction band. The different species can be particles (represented by creation operators) or holes (represented by annihilation operators). Since we have translational symmetry, states can be labelled by total momentum $\bs{p}$. We can write the total momentum in terms of the momentum of each particle/hole in the bound state as $\bs{K}\equiv \sum \nu_i \bs{k}_i$ where $\nu_i$ is $+1$ if the $i^{\mathrm{th}}$ operator is a creation operator and $-1$ if it is an annihilation operator. For a bound state with total momentum $\bs{p}$, we only sum over sets of momenta $(\bs{k})$ which have total momentum $\bs{K} = \bs{p}$. The ground state is $\ket{\mathrm{GS}} = \prod_{k, j} c^\dagger_{\bs{k}, v_j} \ket{0}$, where $c^\dagger_{\bs{k}, v_j}$ creates an electron in the $j^{\mathrm{th}}$ valence band.

We now show that the normalisation is $\mathcal{N} = \frac{1}{\sqrt{\prod_\alpha N_\alpha!}}$ when the bound state wave function is normalised to $\sum_{\bs{K} = \bs{p}} |\phi^{(\bs{k})}|^2=1$. 

We first study a trion example before generalising to arbitrary bound states. We consider a trion with two electrons in the conduction band $c$ and a hole in the valence band $v$. This has wave function,
\begin{equation}
\ket{\bs{p}_{\mathrm{trion}}}= \mathcal{
N}\sum_{\bs{k}+\bs{q}-\bs{l} = \bs{p}} \phi^{(\bs{k},\bs{q},\bs{l})}c^\dagger_{\bs{k}, c}c^\dagger_{\bs{q}, c}c_{\bs{l}, v}\ket{\mathrm{GS}}
\end{equation}
with $\ket{\mathrm{GS}} = \prod_{\bs{k}} c^\dagger_{\bs{k}, v}\ket{0}$. We calculate,
\begin{align}
\braket{\bs{p}_{\mathrm{trion}}|\bs{p}_{\mathrm{trion}}} &= |\mathcal{
N}|^2 \sum_{\bs{k}+\bs{q}-\bs{l} = \bs{p}} \sum_{\bs{k}'+\bs{q}'-\bs{l}' = \bs{p}} \bar\phi^{(\bs{k}',\bs{q}',\bs{l}')}\phi^{(\bs{k},\bs{q},\bs{l})} \bra{\mathrm{GS}}c^\dagger_{\bs{l}', v}c_{\bs{q}', c}c_{\bs{k}', c}c^\dagger_{\bs{k}, c}c^\dagger_{\bs{q}, c}c_{\bs{l}, v}\ket{\mathrm{GS}}\\
&= |\mathcal{
N}|^2 \sum_{\bs{k}+\bs{q}-\bs{l} = \bs{p}} \sum_{\bs{k}'+\bs{q}'-\bs{l}' = \bs{p}} \bar\phi^{(\bs{k}',\bs{q}',\bs{l}')}\phi^{(\bs{k},\bs{q},\bs{l})} \bra{\mathrm{GS}}c_{\bs{q}', c}c_{\bs{k}', c}c^\dagger_{\bs{k}, c}c^\dagger_{\bs{q}, c}\ket{\mathrm{GS}}\bra{\mathrm{GS}}c^\dagger_{\bs{l}', v}c_{\bs{l}, v}\ket{\mathrm{GS}}\\
&= |\mathcal{
N}|^2 \sum_{\bs{k}+\bs{q}-\bs{l} = \bs{p}} \sum_{\bs{k}'+\bs{q}'-\bs{l}' = \bs{p}} \bar\phi^{(\bs{k}',\bs{q}',\bs{l}')}\phi^{(\bs{k},\bs{q},\bs{l})} (\delta_{\bs{k}, \bs{k}'}\delta_{\bs{q}, \bs{q}'} - \delta_{\bs{q}', \bs{k}}\delta_{\bs{k}', \bs{q}})\delta_{\bs{l}, \bs{l}'}\\
&= |\mathcal{
N}|^2 \sum_{\bs{k}+\bs{q}-\bs{l} = \bs{p}} |\phi^{(\bs{k},\bs{q},\bs{l})}|^2 - \bar\phi^{(\bs{q},\bs{k},\bs{l})}\phi^{(\bs{k},\bs{q},\bs{l})}.
\end{align}
Now we use the fact that the envelope function is antisymmetric $\phi^{(\bs{k},\bs{q},\bs{l})} = -\phi^{(\bs{q}, \bs{k}, \bs{l})}$.
\begin{align}
\braket{\bs{p}_{\mathrm{trion}}|\bs{p}_{\mathrm{trion}}}&= 2|\mathcal{
N}|^2 \sum_{\bs{k}+\bs{q}-\bs{l} = \bs{p}} |\phi^{(\bs{k},\bs{q},\bs{l})}|^2.
\end{align}
Therefore, if we choose $\sum_{\bs{k}+\bs{q}-\bs{l} = \bs{p}} |\phi^{(\bs{k},\bs{q},\bs{l})}|^2 = 1$ then for $\braket{\bs{p}_{\mathrm{trion}}|\bs{p}_{\mathrm{trion}}} = 1$ we need $\mathcal{
N} = \frac{1}{\sqrt{2}}$.

We now generalise this method to arbitrary bound states,
\begin{align}
\braket{\bs{p}|\bs{p}} &= |\mathcal{
N}|^2\sum_{\bs{K} = \bs{p}} \sum_{\bs{Q} = \bs{p}} \bar \phi^{(\bs{q})}\phi^{(\bs{k})}\\& \quad\quad\bra{\mathrm{GS}}c^\dagger_{\bs{q}_N, \alpha_N}c^\dagger_{\bs{q}_{N-1}, \alpha_{N-1}}\dots c_{\bs{q}_{2}, \alpha_{2}}c_{\bs{q}_{1}, \alpha_{1}}c^\dagger_{\bs{k}_1, \alpha_1}c^\dagger_{\bs{k}_2, \alpha_2}\dots c_{\bs{k}_{N-1}, \alpha_{N-1}}c_{\bs{k}_{N}, \alpha_{N}}\ket{\mathrm{GS}}\nonumber\\
&=
|\mathcal{N}|^2 \sum_{\bs{K} = \bs{p}} \sum_{\bs{Q} = \bs{p}} \bar{\phi}^{(\bs{q})} \phi^{(\bs{k})} \cdot \prod_{\alpha\in \mathrm{particles}} \bra{\mathrm{GS}} \left( \prod_{j: \alpha_j = \alpha}^{\leftarrow} c_{\bs{q}_j, \alpha} \right) \left( \prod_{i: \alpha_i = \alpha}^{\rightarrow} c^\dagger_{\bs{k}_i, \alpha} \right) \ket{\mathrm{GS}} \\&\nonumber\qquad\qquad\qquad\qquad\qquad\qquad\qquad\quad\cdot \prod_{\beta\in \mathrm{holes}} \bra{\mathrm{GS}} \left( \prod_{j: \alpha_j = \beta}^{\leftarrow} c^\dagger_{\bs{q}_j, \beta} \right) \left( \prod_{i: \alpha_i = \beta}^{\rightarrow} c_{\bs{k}_i, \beta} \right) \ket{\mathrm{GS}}.\label{eq:normalisationExpec}
\end{align}
The arrows indicate the relative ordering of operators within each species, which is needed
to avoid incorrect sign factors. The product over $\alpha$ is over all distinguishable species that are particles \emph{i.e.} creation operators on top of the ground state. The product over $\beta$ is over hole species. We consider the expectation value over just particle species $\alpha$. This can be simplified using Wick's theorem to give,
\begin{align}
\bra{\mathrm{GS}} \left( \prod_{j: \alpha_j = \alpha}^{\leftarrow} c_{\bs{q}_j, \alpha} \right) \left( \prod_{i: \alpha_i = \alpha}^{\rightarrow} c^\dagger_{\bs{k}_i, \alpha} \right) \ket{\mathrm{GS}} = \sum_{\sigma_\alpha \in S_{N_\alpha}}{\mathrm{sgn}(\sigma_\alpha)}  
     \prod_{i: \alpha_i = \alpha} \delta_{q_i, k_{\sigma_\alpha(i)}},
\end{align}
where we sum over all permutations in the permutation group $S_{N_\alpha}$ where $N_\alpha$ is the number of particles of species $\alpha$. Each permutation has a signature (or sign) $\mathrm{sgn(}\sigma_\alpha)\in \{\pm 1\}$ which ensures antisymmetry. The signature (or sign) of a permutation is $+1$ if the permutation can be decomposed into an even number of transpositions (swaps of two elements), and $-1$ if it requires an odd number. The elements of the permutation group are $\sigma_\alpha$ which has components $\sigma_\alpha(i)$ that give the permuted index. The number of permutations in the group $S_{N_\alpha}$ is $N_\alpha!$. We get exactly the same expression for a hole species. 

We now plug this into Eq.~\eqref{eq:normalisationExpec} to give,
\begin{align}
\braket{\bs{p}|\bs{p}}&=
|\mathcal{N}|^2 \sum_{\bs{K} = \bs{p}} \sum_{\bs{Q} = \bs{p}} \bar{\phi}^{(\bs{q})} \phi^{(\bs{k})} \cdot \prod_{\alpha} \sum_{\sigma_\alpha \in S_{N_\alpha}}{\mathrm{sgn}(\sigma_\alpha)}  
     \prod_{i: \alpha_i = \alpha} \delta_{q_i, k_{\sigma_\alpha(i)}}.\label{eq:normalisationSimplification}
\end{align}
The product is now over all species --- particles and holes. The bound state wave function is already anti-symmetric so that,
\begin{equation}
\left(\phi^{(\bs{k}_{\sigma_{\alpha_1}(1)},\dots,\bs{k}_{\sigma_{\alpha_N}(N)})}\right)\phi^{(\bs{k}_{1},\dots,\bs{k}_{N})} \prod_\beta\mathrm{sgn}(\sigma_\beta)= |\phi^{(\bs{k}_{1},\dots,\bs{k}_{N})} |^2.
\end{equation}
This means each permutation will give the same contribution to $\braket{\bs{p}|\bs{p}}$. There are $N_\alpha!$ terms in the permutation of a given species $\alpha$ and so the total product over all species $\alpha$ in Eq.~\eqref{eq:normalisationSimplification} will give a factor of $\prod_\alpha (N_\alpha !)$. Therefore we find that,
\begin{align}
\braket{\bs{p}|\bs{p}}&=
|\mathcal{N}|^2 \sum_{\bs{K} = \bs{p}}   |\phi^{(\bs{k})}|^2 \prod_{\alpha} N_\alpha!.
\end{align}
For normalisation we want $\braket{\bs{p}|\bs{p}} = 1$. Therefore,
\begin{equation}
\mathcal{N} = \frac{1}{\sqrt{\prod_{\alpha} N_\alpha!}},
\end{equation}
if 
\begin{equation}
\sum_{\bs{K} = \bs{p}} |\phi^{(\bs{k})}|^2=1 .
\end{equation}

\newpage
\section{Bound state Berry connections}\label{sec:BoundStateBerry}
In this section we derive the form of the Wilson loops, Wannier states and Berry connections for arbitrary bound states. We do this in 1D but the calculation can be easily extended to higher dimensions. This calculation is exactly analogous to Ref.~\cite{ExcitonBerryology}. We use the projected position operator for each distinct species of particle and hole in the bound state. Given the bound state wave functions in  Eq.~\eqref{eq:AppboundstateWavefunction} for a given band we can form the projector onto the excitation/bound state band,
\begin{equation}
\hat P_{\mathrm{exc}} = \sum_{ {p}\in\mathrm{BZ}} \ket{ {p}} \bra{ {p}}.
\end{equation}
\subsection{Projected position operator for particles}
If species $\alpha$ are particles (not holes) then we use the number operator projected to the $\alpha$ species, $\hat n_{ {R}}^\alpha = \sum_{ {k},  {k}', i} \bar\psi^{ {k}, \alpha}_{ {R}, i} \psi^{ {k}', \alpha}_{ {R}, i} c^\dagger_{ {k}, \alpha} c_{ {k}', \alpha}$. We have defined the electronic Bloch functions $\psi^{\bs{k}, \beta}_{\bs{R}, i}\equiv e^{\mathrm{i}\bs{k}\cdot \bs{R}} u^{\bs{k}, \beta}_{i} /\sqrt{\mathcal{V}}$ (labelling the unit cells by $\bs{R}$, orbitals within the unit cell by $i$ and species/non-interacting bands by $\beta$, while the normalisation factor $\mathcal{V}$ equals the total number of unit cells). Since in periodic systems the normal position operator is ill-defined, we use the \emph{periodic} position operator~\cite{ExcitonBerryology}. This gives the average (periodic) position of all particles of species $\alpha$, given there are $N_\alpha$ of these this is,
\begin{align}
\hat Z^{\alpha} &= \frac{1}{N_\alpha}\sum_{ {R}} e^{\mathrm{i} \Delta\cdot {R}} \hat n_{ {R}}^\alpha\\
&=  \frac{1}{N_\alpha}\frac{1}{\mathcal{V}}\sum_{ {k},  {k}',  {R}, i} e^{\mathrm{i}( \Delta - {k}+ {k}')\cdot {R}}\:\bar u^{ {k}, \alpha}_{i} u^{ {k}', \alpha}_{i} c^\dagger_{ {k}, \alpha} c_{ {k}', \alpha}
\\
&=  \frac{1}{N_\alpha}\sum_{ {k}, i}\:\bar u^{ {k}+ \Delta, \alpha}_{i} u^{ {k}, \alpha}_{i} c^\dagger_{ {k}+ \Delta, \alpha} c_{ {k}, \alpha}\\
&=  \frac{1}{N_\alpha}\sum_{ {k}}\:\braket{u^{ {k}+ \Delta, \alpha}| u^{ {k}, \alpha} }c^\dagger_{ {k}+ \Delta, \alpha} c_{ {k}, \alpha}
\end{align}
where $ \Delta = \frac{2\pi}{L}$ for system size $L$. We now take the projected position operator,
\begin{align}
\hat{\mathcal{Z}}^{\alpha} &= \hat P_{\mathrm{exc}}\hat Z^{\alpha}\hat P_{\mathrm{exc}}\\
&= \sum_{ {p},  {q}} [\mathcal{Z}^{\alpha}]_{ {q},  {p}} \ket{ {q}}\bra{ {p}},
\end{align}
where,
\begin{equation}
[\mathcal{Z}^{\alpha}]_{ {q},  {p}}  \equiv \bra{ {q}}\hat{Z}^{\alpha} \ket{ {p}}.
\end{equation}
To calculate these matrix elements we use,
\begin{equation}
\hat Z^{\alpha} c^\dagger_{ {k}, \beta} = c^\dagger_{ {k}, \beta} \hat Z^{\alpha} + \frac{1}{N_\alpha}\delta_{\alpha, \beta} \braket{u^{ {k}+ \Delta, \alpha}| u^{ {k}, \alpha}}  c^\dagger_{ {k}+ {\Delta}, \alpha}
\end{equation}
Therefore,
\begin{align}
\hat Z^{\alpha} \ket{ {p}} &= \frac{1}{N_\alpha}\mathcal{N}\sum _{n:\alpha_n = \alpha}\bigg[\sum_{K =  {p}} \phi^{( {k})}\braket{u^{ {k}_n+ \Delta, \alpha}| u^{ {k}_n, \alpha}}\nonumber \\ &\quad\quad\quad\quad\quad\quad \cdot c^\dagger_{ {k}_1, \alpha_1}\dots c^\dagger_{ {k}_n+ {\Delta}, \alpha_n}\dots c_{ {k}_{N}, \alpha_{N}}\ket{\mathrm{GS}}\bigg].
\end{align}
The summation $\sum _{n:\alpha_n = \alpha}$ is over all particles in the bound state which are of species $\alpha$. The particle labelled by $n$ has momentum labelled $ {k}_n$. The total momentum of the resulting state ($\hat Z^{\alpha}_x \ket{ {p}}$) is $ {p}+ {\Delta}$. For many body states, the states at different momenta are orthogonal. This means that when we calculate the expectation value $\bra{ {p}} \hat Z^{\alpha} \ket{ {q}}$, the only non-zero terms are 
\begin{align}
\bra{ {p}+ {\Delta}}\hat Z^{\alpha}  \ket{ {p}} &= \frac{1}{N_\alpha}|\mathcal{N}|^2
\sum _{n:\alpha_n = \alpha}\bigg[\sum_{Q =  {p}+ {\Delta}}\sum_{K =  {p}} \bar \phi^{( {q})} \phi^{({k})}\braket{u^{ {k}_n+ \Delta, \alpha}| u^{ {k}_n, \alpha}} \\*&\quad\quad\quad\quad\quad\quad \cdot  \bra{\mathrm{GS}}c^\dagger_{ {q}_{N}, \alpha_{N}} \dots c_{ {q}_n, \alpha_n}\dots  c_{ {q}_1, \alpha_1} c^\dagger_{ {k}_1, \alpha_1}\dots c^\dagger_{ {k}_n+ {\Delta}, \alpha_n}\dots c_{ {k}_{N}, \alpha_{N}}\ket{\mathrm{GS}}\bigg]\nonumber\\*
&=
\frac{1}{N_\alpha}|\mathcal{N}|^2 \sum _{n:\alpha_n = \alpha}\sum_{K = p} \sum_{Q = p+\Delta} \bar{\phi}^{(q)} \phi^{(k)} \braket{u^{ {k}_n+ \Delta, \alpha}| u^{ {k}_n, \alpha}}\\*&\quad\nonumber\qquad\qquad\qquad\qquad\cdot \prod_{\beta\in \mathrm{particles}} \bra{\mathrm{GS}} \left( \prod_{j: \alpha_j = \beta}^{\leftarrow} c_{q_j, \beta} \right) \left( \prod_{i: \alpha_i = \beta}^{\rightarrow} c^\dagger_{k_i+\delta_{i, n}\Delta 
, \beta} \right) \ket{\mathrm{GS}} \\&\quad\nonumber\qquad\qquad\qquad\qquad\cdot \prod_{\gamma\in \mathrm{holes}} \bra{\mathrm{GS}} \left( \prod_{j: \alpha_j = \gamma}^{\leftarrow} c^\dagger_{q_j, \gamma} \right) \left( \prod_{i: \alpha_i = \gamma}^{\rightarrow} c_{k_i, \gamma} \right) \ket{\mathrm{GS}}\\
&=
\frac{1}{N_\alpha}|\mathcal{N}|^2 \sum _{n:\alpha_n = \alpha}\sum_{K = p} \sum_{Q = p+\Delta} \bar{\phi}^{(q)} \phi^{(k)} \braket{u^{ {k}_n+ \Delta, \alpha}| u^{ {k}_n, \alpha}} \\*&\quad\nonumber\qquad\qquad\qquad\qquad\cdot \prod_{\beta} \sum_{\sigma_\beta \in S_{N_\beta}}{\mathrm{sgn}(\sigma_\beta)}  
     \prod_{i: \alpha_i = \beta} \delta_{q_i, [k_{\sigma_\beta(i)}+\delta_{n, \sigma_\beta(i)}\Delta]}. \nonumber
\end{align}
See Appendix~\ref{sec:Normalisation} for explanation of above notation.

Using the anti-symmetry of $\phi^{(k)}$ we identify,
\begin{equation}
\left(\bar\phi^{(k_{\sigma_{\alpha_1}(1)},\dots, k_{n}+\Delta,\dots,k_{\sigma_{\alpha_N}(N)})}\right)\phi^{(k_{1}, \dots, k_{N})} \prod_\beta\mathrm{sgn}(\sigma_\beta)= \bar \phi^{(k_{1},\dots,k_{n}+\Delta, \dots, k_{N})}\phi^{(k_{1}, \dots, k_n, \dots, k_{N})}.
\end{equation}
Therefore we can simplify the matrix element to,
\begin{align}
\bra{ {p}+ {\Delta}}\hat Z^{\alpha}  \ket{ {p}}&=
\frac{1}{N_\alpha} \sum _{n:\alpha_n = \alpha}\sum_{K = p}  \bar{\phi}^{(k_1, \dots, k_n + \Delta, \dots, k_N)} \phi^{(k_1, \dots, k_n, \dots, k_N)} \braket{u^{ {k}_n+ \Delta, \alpha}| u^{ {k}_n, \alpha}}.
\end{align}
We do one last simplification by identifying that each term in the sum over $n$ is identical since the particles we are summing over are identical. Therefore we just choose one of the particles which are of species $\alpha$. Assuming the species of the $n^{\mathrm{th}}$ particle is $\alpha$ \emph{i.e.} $\alpha_n = \alpha$ we therefore we obtain,
\begin{align}
\bra{ {p}+ {\Delta}}\hat Z^{\alpha}  \ket{ {p}}&=
\sum_{K = p}  \bar{\phi}^{(k_1, \dots, k_n + \Delta, \dots, k_N)} \phi^{(k_1, \dots, k_n, \dots, k_N)} \braket{u^{ {k}_n+ \Delta, \alpha}| u^{ {k}_n, \alpha}}\\
&\equiv t^{\alpha}_{ {p}}.
\end{align}

We therefore can write,
\begin{align}
\hat{\mathcal{Z}}^{\alpha} 
&= \sum_{ {p},  {q}} t^{\alpha}_{ {p}} \ket{ {p}+ {\Delta}}\bra{ {p}}.
\end{align}
We now show that the eigenvectors of $\hat{\mathcal{Z}}^\alpha$ are the maximally localised Wannier states and the eigenvalues give the Wannier centres. We label the $R^{th}$ Wannier state as,
\begin{equation}
\ket{\mathcal{W}_R^\alpha} = \sum_{p} \mathcal{W}_R^\alpha({p}) \ket{p},\label{eq:eigenstateElec}
\end{equation}
which has associated eigenvalue $\lambda^\alpha_R$,
\begin{align}
\hat{\mathcal{Z}} \ket{\mathcal{W}_R^\alpha} &= \sum_p \mathcal{W}_R^\alpha(p)  \hat{\mathcal{Z}}^\alpha\ket{p} \\
&=\sum_p \mathcal{W}_R^\alpha(p) t^\alpha_p \ket{p+\Delta}\\
&=\sum_p \mathcal{W}_R^\alpha({p-\Delta}) t^\alpha_{p-\Delta} \ket{p}.\label{eq:ZonW}
\end{align}
In addition since $\ket{\mathcal{W}_R^\alpha}$ is an eigenvector of $\hat{\mathcal{Z}}^\alpha$,
\begin{align}
\hat{\mathcal{Z}}^\alpha \ket{\mathcal{W}_R^\alpha} &= \lambda^\alpha_R \ket{\mathcal{W}_R^\alpha}\\
&= \sum_p \mathcal{W}_R^\alpha(p) \lambda^\alpha_R \ket{p}.~\label{eq:eigvalueWannierstateeq}
\end{align}
We solve for the components of the eigenvector ($\ket{\mathcal{W}_R^\alpha}$) by equating  Eq.~\eqref{eq:eigvalueWannierstateeq} and Eq.~\eqref{eq:ZonW}. We first find the relationship between $\mathcal{W}_R^\alpha({p-\Delta})$ and $\mathcal{W}_R^\alpha(p)$,
\begin{align}
\mathcal{W}_R^\alpha(p - \Delta) t^\alpha_{p-\Delta} &= \mathcal{W}_R^\alpha(p)  \lambda^\alpha_R,
\end{align}
therefore,
\begin{align}
\mathcal{W}_R^\alpha(p+\Delta)  &=\frac{1}{\lambda^\alpha_R} t^\alpha_{p}\mathcal{W}_R^\alpha(p).
\end{align}
Component $\mathcal{W}_R^\alpha(p+2\Delta)$ is related to $\mathcal{W}_R^\alpha(p)$ by,
\begin{align}
\mathcal{W}_R^\alpha(p+2\Delta)  &=\frac{1}{(\lambda^\alpha_R)^2} t^\alpha_{p+\Delta}t^\alpha_{p}\mathcal{W}_R^\alpha(p).\label{eq:WannierStart}
\end{align}
We can generalise this to give the relationship between arbitrary components $\mathcal{W}^\alpha_R(q)$ and $\mathcal{W}^\alpha_R(p)$,
\begin{align}
\mathcal{W}^\alpha_R(p) &=\frac{1}{(\lambda^\alpha_R)^{(p-q)/\Delta}} \left(\prod_{k = p-\Delta}^q t^\alpha_{k}\right)\mathcal{W}^\alpha_R(q).\label{eq:generalcomponentWannier}
\end{align}
Finally, we find an expression for the eigenvalues ($\lambda_R$) by choosing $p = q+2\pi$, 
\begin{align}
\mathcal{W}^\alpha_R({q+2\pi})  &= \frac{1}{(\lambda^\alpha_R)^{L}} \left(\prod_{k = q+2\pi - \Delta}^q t^\alpha_{k}\right)\mathcal{W}^\alpha_R(q),
\end{align}
and using $\mathcal{W}^\alpha_R({q+2\pi}) = \mathcal{W}^\alpha_R(q)$. This means that,
\begin{equation}
 \frac{1}{(\lambda^\alpha_R)^L}\prod_{k = q+2\pi - \Delta}^q t^\alpha_{k} = 1.
\end{equation}
It follows that,
\begin{align}
(\lambda^\alpha_R)^L &=  \prod_{k = q+2\pi - \Delta}^q t^\alpha_{k}\\
&=  \prod_{k = 0}^{2\pi - \Delta} t^\alpha_{k}\\
&\equiv W^\alpha. \label{eq:WilsonLoopDef}
\end{align}
Where we have identified $W^\alpha$ as the Wilson loop for species $\alpha$. Therefore the eigenvalues are,
\begin{equation}
\lambda^\alpha_R = (W^\alpha)^{1/L} e^{\mathrm{i}\Delta R}.\label{eq:EigvalueFinite}
\end{equation}

We now show that the eigenstates are the Wannier states. We begin with Eq.~\eqref{eq:generalcomponentWannier} and plug in the eigenvalue expression in Eq.~\eqref{eq:EigvalueFinite},
\begin{align}
\mathcal{W}^\alpha_R(q) &=\frac{1}{(\lambda^\alpha_R)^{(q-p)/\Delta}} \left(\prod_{k = q-\Delta}^p t^\alpha_{k}\right)\mathcal{W}^\alpha_R(p)\\
&=\frac{1}{\left[(W^\alpha)^{\frac{1}{L}} e^{\mathrm{i}\Delta R}\right]^{(q-p)/\Delta}} \left(\prod_{k = q-\Delta}^p t^\alpha_{k}\right)\mathcal{W}^\alpha_R(p)\\
&= e^{-\mathrm{i}(q-p)R} (W^\alpha)^{-\frac{q - p}{2\pi}}
 \left(\prod_{k = q-\Delta}^p t^\alpha_{k}\right)\mathcal{W}^\alpha_R(p).
\end{align}
If we set $p = 0$ then we can construct the Wannier states using Eq.~\eqref{eq:eigenstateElec},
\begin{equation}
\ket{\mathcal{W}^\alpha_R} = \mathcal{W}^\alpha_R(0) \sum_{p} e^{-\mathrm{i} pR} (W^\alpha)^{-\frac{p}{2\pi}} \prod_{k = 0}^{p-\Delta} t^\alpha_{k}\ket{p}.
\end{equation}
We rewrite this as,
\begin{equation}
\ket{\mathcal{W}^\alpha_R} = \mathcal{N} \sum_{p} e^{-\mathrm{i} pR} (W^\alpha)^{-\frac{p}{2\pi}} \prod_{k = 0}^{p-\Delta} t^\alpha_{k}\ket{p},\label{eq:WannierStatesFinite}
\end{equation}
where we have rewritten $\mathcal{W}^\alpha_R(0)$ as some normalisation $\mathcal{N}$. We see that this takes the form of Wannier states \emph{i.e.} the inverse Fourier transform of the states within the band $\ket{p}$. However, in addition we have some prefactor in front of each $\ket{p}$ which modifies its phase. This ensures the resulting Wannier states are maximally localised. 

Now we take the thermodynamic limit of the Wilson loop $W^\alpha$, in order to extract the form of the Berry connection for the $\alpha^{\mathrm{th}}$ species. We know in general that the Wilson loop has modulus 1 in the thermodynamic limit and goes as $e^{\mathrm{i}\int_0^{2\pi} \mathrm{d}k\:A^\alpha(k) }$ where $A^\alpha(k)$ is the Berry connection for the $\alpha^{\mathrm{th}}$ species. Taking the thermodynamic limit of Eq.~\eqref{eq:WilsonLoopDef} we obtain,
\begin{align}
W^\alpha &=  \prod_{p = 0}^{2\pi - \Delta} t^\alpha_{p}\\
&=\prod_{p = 0}^{2\pi - \Delta} \left[\sum_{K =  {p}}\bar \phi^{(\dots,  {k}_n+ {\Delta}, \dots)}\phi^{(\dots,  {k}_n, \dots)}\braket{u^{ {k}_n+ \Delta, \alpha}| u^{ {k}_n, \alpha}}\right]\\
&=\prod_{p = 0}^{2\pi - \Delta} \left[\sum_{K =  {p}}\left(\bar \phi^{(\dots,  {k}_n, \dots)}+ {\Delta} \partial_{k_n}\bar \phi^{(\dots,  {k}_n, \dots)}\right)\phi^{(\dots,  {k}_n, \dots)}\left(\bra{u^{ {k}_n, \alpha}}+ \Delta \partial_{k_n}\bra{u^{ {k}_n, \alpha} }\right) \ket{u^{ {k}_n, \alpha}}\right].
\end{align}
For clarity of notation we return to our previous more compact notation $\phi^{(\dots,  {k}_n, \dots)} = \phi^{(k)}$,
\begin{align}
W^\alpha&=\prod_{p = 0}^{2\pi - \Delta} \left\{1+\Delta\sum_{K = p}\left[\partial_{k_n}\left(\bar \phi^{(k)}\right) \phi^{(k)} +|\phi^{(k)}|^2\partial_{k_n}\left(\bra{u^{ {k}_n, \alpha} }\right) \ket{u^{ {k}_n, \alpha}}\right] +\mathcal{O}(\Delta^2)\right\}\\
&=\prod_{p = 0}^{2\pi - \Delta} \left[1-\Delta\sum_{K = p}\left( \bar \phi^{(k)} \partial_{k_n}\phi^{(k)} + |\phi^{(k)}|^2\bra{u^{ {k}_n, \alpha} }\partial_{k_n}\ket{u^{ {k}_n, \alpha}}\right) +\mathcal{O}(\Delta^2)\right]\\
&=\prod_{p = 0}^{2\pi - \Delta} \left[1+\mathrm{i}\Delta \mathcal{A}^{\alpha}(p) +\mathcal{O}(\Delta^2)\right]\\
&\approx \prod_{p = 0}^{2\pi - \Delta} \left[e^{\mathrm{i}\Delta \mathcal{A}^{\alpha}(p)}\right]\\
&\approx  e^{\mathrm{i}\int_{0}^{2\pi}\mathrm{d}p\: \mathcal{A}^{\alpha}(p)},
\end{align}
where the $\approx$ becomes an equality in the thermodynamic limit. We have defined the Berry connection for the $\alpha^{\mathrm{th}}$ species as, 
\begin{equation}
\mathcal{A}^\alpha (p) \equiv \sum_{K = p}\left( \bar \phi^{(k)}\mathrm{i} \partial_{k_n}\phi^{(k)} + |\phi^{(k)}|^2\bra{u^{ {k}_n, \alpha} }\mathrm{i}\partial_{k_n}\ket{u^{ {k}_n, \alpha}}\right). 
\end{equation}
We can generalise this to arbitrary dimensions to get,
\begin{align}
\bs{\mathcal{A}}^\alpha (p) &\equiv \sum_{\bs{K} =  {\bs{p}}}\left( \bar \phi^{(\bs{k})}\mathrm{i} \nabla_{\bs{k}_n}\phi^{(\bs{k})} + |\phi^{(\bs{k})}|^2\bra{u^{ {\bs{k}}_n, \alpha} }\mathrm{i}\nabla_{\bs{k}_n}\ket{u^{ {\bs{k}}_n, \alpha}}\right)\\
&\equiv \sum_{\bs{K} =  {\bs{p}}}\left( \bar \phi^{(\bs{k})}\mathrm{i} \nabla_{\bs{k}_n}\phi^{(\bs{k})} + |\phi^{(\bs{k})}|^2\bs{A}_{\mathrm{elec}, \alpha}(\bs{k}_n)\right)
\end{align}
where $\bs{A}_{\mathrm{elec}, \alpha}(\bs{k}_n) = \bra{u^{ {\bs{k}}_n, \alpha} }\mathrm{i}\nabla_{\bs{k}_n}\ket{u^{ {\bs{k}}_n, \alpha}}$. For calculations however it is often easier to use the gauge invariant expression for the Wilson loop,
\begin{align}
    W^\alpha&=\prod_{p = 0}^{2\pi - \Delta} \left[\sum_{K = p}\bar \phi^{(\dots,  {k}_n+ {\Delta}, \dots)}\phi^{(\dots,  {k}_n, \dots)}\braket{u^{ {k}_n+ \Delta, \alpha}| u^{ {k}_n, \alpha}}\right].\label{eq:WilsonLoop}
\end{align}

\subsection{Projected position operator for holes}
We now repeat the calculation but for hole excitations. We use the projected position operator for holes. This is,
\begin{align}
\hat Z^{\alpha} &= \frac{1}{N_\alpha}\sum_{ {R}} e^{\mathrm{i} \Delta\cdot {R}} \hat n_{ {R}}^\alpha\\
&=  \frac{1}{N_\alpha}\frac{1}{\mathcal{V}}\sum_{ {k},  {k}',  {R}, i} e^{\mathrm{i}( \Delta + {k}- {k}')\cdot {R}}\: u^{ {k}, \alpha}_{i} \bar u^{ {k}', \alpha}_{i} c_{ {k}, \alpha} c^\dagger_{ {k}', \alpha}
\\
&=  \frac{1}{N_\alpha}\sum_{ {k}, i}\:u^{ {k}, \alpha}_{i} \bar u^{ {k}+ \Delta, \alpha}_{i} c_{ {k}, \alpha} c^\dagger_{ {k}+\Delta, \alpha}\\
&=  \frac{1}{N_\alpha}\sum_{ {k}}\:\braket{u^{ {k}, \alpha}| u^{ {k}-\Delta, \alpha} }c_{ {k}- \Delta, \alpha} c^\dagger_{ {k}, \alpha}.
\end{align}
We again use the operator,
\begin{align}
\hat{\mathcal{Z}}^{\alpha} &= \hat P_{\mathrm{exc}}\hat Z^{\alpha}\hat P_{\mathrm{exc}}\\
&= \sum_{ {p},  {q}} [\mathcal{Z}^{\alpha}]_{ {q},  {p}} \ket{ {q}}\bra{ {p}},
\end{align}
where,
\begin{equation}
[\mathcal{Z}^{\alpha}]_{ {q},  {p}}  \equiv \bra{ {q}}\hat{Z}^{\alpha} \ket{ {p}}.
\end{equation}
To calculate these matrix elements we use,
\begin{equation}
\hat Z^{\alpha} c_{ {k}, \beta} = c_{ {k}, \beta} \hat Z^{\alpha} + \frac{1}{N_\alpha}\delta_{\alpha, \beta} \braket{u^{ {k}, \alpha}| u^{ {k}-\Delta, \alpha}}  c_{ {k}- {\Delta}, \alpha}
\end{equation}
Therefore,
\begin{align}
\hat Z^{\alpha} \ket{ {p}} &= \frac{1}{N_\alpha}\mathcal{N}\sum _{n:\alpha_n = \alpha}\bigg[\sum_{K = p} \phi^{( {k})}\braket{u^{ {k}_n, \alpha}| u^{ {k}_n- \Delta, \alpha}}\nonumber \\ &\quad\quad\quad\quad\quad\quad \cdot c^\dagger_{ {k}_1, \alpha_1}\dots c_{ {k}_n- {\Delta}, \alpha_n}\dots c_{ {k}_{N}, \alpha_{N}}\ket{\mathrm{GS}}\bigg].
\end{align}
Therefore we can simplify the matrix element to,
\begin{align}
\bra{ {p}+ {\Delta}}\hat Z^{\alpha}  \ket{ {p}}&=
\sum_{K = p}  \bar{\phi}^{(k_1, \dots, k_n - \Delta, \dots, k_N)} \phi^{(k_1, \dots, k_n, \dots, k_N)} \braket{u^{ {k}_n, \alpha}| u^{ {k}_n-\Delta, \alpha}}\\
&\equiv t_{p}^{\alpha}
\end{align}
The Wilson loop is,
\begin{align}
W^\alpha &=  \prod_{p = 0}^{2\pi - \Delta} t^\alpha_{p}\\
&=\prod_{p = 0}^{2\pi - \Delta} \left[\sum_{K = p}  \bar{\phi}^{(k_1, \dots, k_n - \Delta, \dots, k_N)} \phi^{(k_1, \dots, k_n, \dots, k_N)} \braket{u^{ {k}_n, \alpha}| u^{ {k}_n-\Delta, \alpha}}\right]\\
&=\prod_{p = 0}^{2\pi - \Delta} \left[\sum_{K = p}\left(\bar \phi^{(\dots,  {k}_n, \dots)}- {\Delta} \partial_{k_n}\bar \phi^{(\dots,  {k}_n, \dots)}\right)\phi^{(\dots,  {k}_n, \dots)}\bra{u^{ {k}_n, \alpha}} \left(\ket{u^{ {k}_n, \alpha}}- \Delta \partial_{k_n}\ket{u^{ {k}_n, \alpha} }\right)\right].
\end{align}
Returning to the simplified notation $\phi^{(k)}=\phi^{(\dots,  {k}_n, \dots)}$,
\begin{align}
W^\alpha&=\prod_{p = 0}^{2\pi - \Delta} \left\{1-\Delta\sum_{K = p}\left[\partial_{k_n}\left(\bar \phi^{(k)}\right) \phi^{(k)} + |\phi^{(k)}|^2\bra{u^{ {k}_n, \alpha} } \partial_{k_n}\ket{u^{ {k}_n, \alpha}}\right] +\mathcal{O}(\Delta^2)\right\}\\
&=\prod_{p = 0}^{2\pi - \Delta} \left[1-\Delta\sum_{K =  {p}}\left( -\bar \phi^{(k)} \partial_{k_n}\phi^{(k)} +|\phi^{(k)}|^2\bra{u^{ {k}_n, \alpha} }\partial_{k_n}\ket{u^{ {k}_n, \alpha}}\right) +\mathcal{O}(\Delta^2)\right]\\
&=\prod_{p = 0}^{2\pi - \Delta} \left[1+\mathrm{i}\Delta \mathcal{A}^{\alpha}(p) +\mathcal{O}(\Delta^2)\right]\\
&\approx \prod_{p = 0}^{2\pi - \Delta} \left[e^{\mathrm{i}\Delta \mathcal{A}^{\alpha}(p)}\right]\\
&\approx  e^{\mathrm{i}\int_{0}^{2\pi}\mathrm{d}p\: \mathcal{A}^{\alpha}(p)}.
\end{align}
We have defined the Berry connection for the $\alpha^{\mathrm{th}}$ species as, 
\begin{equation}
\mathcal{A}^\alpha (p) \equiv \sum_{K =  {p}}\left( -\bar \phi^{(k)}\mathrm{i} \partial_{k_n}\phi^{(k)} + |\phi^{(k)}|^2\bra{u^{ {k}_n, \alpha} }\mathrm{i}\partial_{k_n}\ket{u^{ {k}_n, \alpha}}\right). 
\end{equation}
We can generalise this to arbitrary dimensions to get,
\begin{align}
\bs{\mathcal{A}}^\alpha (p) &\equiv \sum_{\bs{K} =  {\bs{p}}}\left(- \bar \phi^{(\bs{k})}\mathrm{i} \nabla_{\bs{k}_n}\phi^{(\bs{k})} + |\phi^{(\bs{k})}|^2\bra{u^{ {\bs{k}}_n, \alpha} }\mathrm{i}\nabla_{\bs{k}_n}\ket{u^{ {\bs{k}}_n, \alpha}}\right)\\
&\equiv \sum_{\bs{K} =  {\bs{p}}}\left( -\bar \phi^{(\bs{k})}\mathrm{i} \nabla_{\bs{k}_n}\phi^{(\bs{k})} + |\phi^{(\bs{k})}|^2\bs{A}_{\mathrm{elec}, \alpha}(\bs{k}_n)\right)
\end{align}
where $\bs{A}_{\mathrm{elec}, \alpha}(\bs{k}_n) = \bra{u^{ {\bs{k}}_n, \alpha} }\mathrm{i}\nabla_{\bs{k}_n}\ket{u^{ {\bs{k}}_n, \alpha}}$. Lastly, we also have the gauge invariant expression for the Wilson loop,
\begin{align}
    W^\alpha&=\prod_{p = 0}^{2\pi - \Delta} \left[\sum_{K = p}  \bar{\phi}^{(k_1, \dots, k_n - \Delta, \dots, k_N)} \phi^{(k_1, \dots, k_n, \dots, k_N)} \braket{u^{ {k}_n, \alpha}| u^{ {k}_n-\Delta, \alpha}}\right].
\end{align}
\newpage
\section{Gauge invariant quantities}
\label{sec:PhysicalMeaningofD}
\subsection{Physical meaning}
In this section we give the physical meaning of the gauge invariant quantities $\bs{\mathcal{D}}^{\alpha, \beta}(\bs{p})$ which are defined as,
\begin{equation}
    \bs{\mathcal{D}}^{\alpha, \beta}(\bs{p}) = \bs{\mathcal{A}}^{\alpha}(\bs{p})-\bs{\mathcal{A}}^{\beta}(\bs{p}).
\end{equation}
We show that this quantity is,
\begin{equation}
\bs{\mathcal{D}}^{\alpha, \beta}(\bs{p}) = \bra{\bs{p}}\left(\hat{\bs{R}}^{\alpha} - \hat{\bs{R}}^{\beta}\right)\ket{\bs{p}},
\end{equation}
recalling the definitions given in the main text. If species $\alpha$ is a particle (not hole) we construct $\hat{\bs{R}}^\alpha$ by dropping terms which depend on other species to give,
\begin{equation}
\hat{\bs{R}}^\alpha = \frac{1}{N_\alpha}\sum_{\bs{k}, \bs{k}', \bs{R}, i} \bs{R} \:\bar\psi^{\bs{k}, \alpha}_{\bs{R}, i} \psi^{\bs{k}', \alpha}_{\bs{R}, i} c^\dagger_{\bs{k}, \alpha} c_{\bs{k}', \alpha},
\end{equation}
where we have divided by the number of particles of species $\alpha$ in the bound state ($N_\alpha$) to obtain an average position. 
For holes we instead define, 
\begin{equation}
\hat{\bs{R}}^\alpha = \frac{1}{N_\alpha}\sum_{\bs{k}, \bs{k}', \bs{R}, i} \bs{R} \:\psi^{\bs{k}, \alpha}_{\bs{R}, i} \bar \psi^{\bs{k}', \alpha}_{\bs{R}, i} c_{\bs{k}, \alpha} c^\dagger_{\bs{k}', \alpha}.
\end{equation}
This means that $\bs{\mathcal{D}}^{\alpha, \beta}(\bs{p})$ is equal to the average distance between the average location of species $\alpha$ particles and species $\beta$. 

To show this we proceed in 1D before generalising to higher dimensions. We use the identity,
\begin{align}
\bra{ {p}+ {\Delta}}\hat Z^{\alpha}  \ket{ {p}}&=
1+ \mathrm{i} \Delta \mathcal{A^{\alpha}} (p) + \mathcal{O}(\Delta^2).
\end{align}
(See proof in Appendix~\ref{sec:BoundStateBerry}).

Since $\hat Z^{\alpha}$ is off diagonal in momentum, we consider a wave packet,
\begin{equation}
\ket{\psi} =\int_{\mathrm{BZ}} {\mathrm{d}p} \:
a_{p}\ket{p}.
\end{equation}
where $\braket{\psi|\psi} =1$ so $\int_{\mathrm{BZ}}|a_p|^2 \mathrm{d}p= 1$.

For bound states the relative position operator $\hat{R}^{\alpha} - \hat R^{\beta}$ becomes well-defined since the bound state is exponentially localised in this coordinate. Therefore, 
\begin{align}
\bra{\psi}\left(\hat{R}^{\alpha} - \hat R^{\beta}\right)\ket{\psi} &= -\lim_{\Delta \rightarrow0}\left[\frac{\mathrm{i}}{\Delta}\int_{\mathrm{BZ}} {\mathrm{d}p} \:
\bar a_{p+\Delta} a_{p} \left(\bra{ {p}+ {\Delta}}\hat Z^{\alpha}  \ket{ {p}} - \bra{ {p}+ {\Delta}}\hat Z^{\beta}  \ket{ {p}}\right)\right]\\
&= \lim_{\Delta \rightarrow0}\left\{\int_{\mathrm{BZ}} {\mathrm{d}p} \:
\bar a_{p+\Delta} a_{p} \left[\mathcal{A}^{\alpha}(p) - \mathcal{A}^{\beta}(p)\right]\right\}\\
&= \lim_{\Delta \rightarrow0}\left\{\int_{\mathrm{BZ}} {\mathrm{d}p} \:
 \left(\bar a_{p}+\Delta \partial_{p} \bar a_{p}\right) a_{p} \left[\mathcal{A}^{\alpha}(p) - \mathcal{A}^{\beta}(p)\right]\right\}\\
 &=\int_{\mathrm{BZ}} {\mathrm{d}p} \:
 |a_{p}|^2 \left[\mathcal{A}^{\alpha}(p) - \mathcal{A}^{\beta}(p)\right]
\end{align}
Taking the limit where we have just one momentum contributing to the integral we have proven that in the thermodynamic limit,
\begin{equation}
\bra{p}\left(\hat{R}^{\alpha} - \hat R^{\beta}\right)\ket{p} = \mathcal{A}^{\alpha}(p) - \mathcal{A}^{\beta}(p)
\end{equation}
We can extend this to higher dimensions to give,
\begin{equation}
\bra{\bs{p}}\left(\hat{\bs{R}}^{\alpha} - \hat{\bs{R}}^{\beta}\right)\ket{\bs{p}} = \bs{\mathcal{A}}^{\alpha}(\bs{p}) - \bs{\mathcal{A}}^{\beta}(\bs{p}).
\end{equation}
Hence we have proven our original statement that,
\begin{equation}
\bs{\mathcal{D}}^{\alpha, \beta}(\bs{p}) = \bra{\bs{p}}\left(\hat{\bs{R}}^{\alpha} - \hat{\bs{R}}^{\beta}\right)\ket{\bs{p}}.
\end{equation}
Since the operators $\hat{\bs{R}}^{\alpha}$ are the average position of particles of species $\alpha$, we can interpret the gauge invariant $\bs{\mathcal{D}}^{\alpha, \beta}(\bs{p})$ as being the average distance between the average position of particles of species $\alpha$ and that of $\beta$.
\subsection{Expression without smooth gauge}
The gauge invariant expression for the QGD depends on whether the two species are both particles, both holes or one of each. 
We can find the component of the QGD along any given direction. Without loss of generality here we consider the $x$ component and so define $\bs{\Delta}_x \equiv \frac{2\pi}{L}\hat{\bs{x}}$ with $L$ the number of unit cells in the $x$ direction.

For one particle, one hole we can write the gauge invariant expression for the QGD as,
\begin{align}
    e^{\mathrm{i}\bs{\Delta}_{x}\cdot \bs{\mathcal{D}}^{\alpha, \beta}(\bs{p})}&= \sum_{\bs{K} = \bs{p}}  \bar{\phi}^{(\bs{k}_1, \dots, \bs{k}_n , \dots, \bs{k}_m\dots, \bs{k}_N)} {\phi}^{(\bs{k}_1, \dots, \bs{k}_n-\bs{\Delta}_x , \dots, \bs{k}_m-\bs{\Delta}_x\dots, \bs{k}_N)}  \braket{u^{ \bs{k}_m, \alpha}| u^{ \bs{k}_m-\bs{\Delta}_x, \alpha}} \braket{u^{ \bs{k}_n - \bs{\Delta}_x, \beta}| u^{ \bs{k}_n, \beta}},
\end{align}
where $n$ labels a hole of species $\beta$ and $m$ labels a particle of species $\alpha$. This can be verified by Taylor expanding the expression in powers of $\bs{\Delta}_x$.

For two particles the expression is,
\begin{align}
    e^{\mathrm{i}\bs{\Delta}_{x}\cdot \bs{\mathcal{D}}^{\alpha, \beta}(\bs{p})}&= \sum_{\bs{K} = \bs{p}}  \bar{\phi}^{(\bs{k}_1, \dots, \bs{k}_n , \dots, \bs{k}_m\dots, \bs{k}_N)} {\phi}^{(\bs{k}_1, \dots, \bs{k}_n+\bs{\Delta}_x , \dots, \bs{k}_m-\bs{\Delta}_x\dots, \bs{k}_N)}  \braket{u^{ \bs{k}_m, \alpha}| u^{ \bs{k}_m-\bs{\Delta}_x, \alpha}} \braket{u^{ \bs{k}_n, \beta}| u^{ \bs{k}_n+\bs{\Delta}_x, \beta}},
\end{align}
where $n$ labels a particle of species $\beta$ and $m$ labels a particle of species $\alpha$.

For two holes the expression is,
\begin{align}
    e^{\mathrm{i}\bs{\Delta}_{x}\cdot \bs{\mathcal{D}}^{\alpha, \beta}(\bs{p})}&= \sum_{\bs{K} = \bs{p}}  \bar{\phi}^{(\bs{k}_1, \dots, \bs{k}_n , \dots, \bs{k}_m\dots, \bs{k}_N)} {\phi}^{(\bs{k}_1, \dots, \bs{k}_n-\bs{\Delta}_x , \dots, \bs{k}_m+\bs{\Delta}_x\dots, \bs{k}_N)}  \braket{u^{ \bs{k}_m+\bs{\Delta}_x, \alpha}| u^{ \bs{k}_m, \alpha}} \braket{u^{ \bs{k}_n-\bs{\Delta}_x, \beta}| u^{ \bs{k}_n, \beta}},
\end{align}
where $n$ labels a particle of species $\beta$ and $m$ labels a particle of species $\alpha$.

\newpage
\section{Electromagnetic gauge transformation}
\label{sec:UnitaryTransformation}
We study the time-dependent Schrodinger equation with some time dependent unitary transformation $\hat U(t)$,
\begin{align}
\mathrm{i} \partial_t \ket{\psi} &= \hat H \ket{\psi}\\
\mathrm{i} \partial_t \ket{\psi} &=\hat H \hat U^\dagger \hat U 
\ket{\psi}\\
\hat U(t)\mathrm{i} \partial_t \ket{\psi} &=\left[\hat U(t)\hat H \hat U^\dagger(t)\right] \hat U(t) \ket{\psi}\\
\mathrm{i} \partial_t \left[\hat U(t)\ket{\psi}\right] -  \mathrm{i} \partial_t \left[\hat U(t)\right]\ket{\psi}&=\left[\hat U(t)\hat H \hat U^\dagger(t)\right] \hat U(t) \ket{\psi}\\
\mathrm{i} \partial_t \left[\hat U(t)\ket{\psi}\right] -  \mathrm{i} \partial_t \left[\hat U(t)\right]\hat U^\dagger(t) \hat U(t)\ket{\psi}&=\left[\hat U(t)\hat H \hat U^\dagger(t)\right] \hat U(t) \ket{\psi}
\end{align}
Therefore,
\begin{equation}
\mathrm{i} \partial_t \left[\hat U(t)\ket{\psi}\right]  =\left[\hat U(t)\hat H \hat U^\dagger(t) + \mathrm{i} \partial_t\left[\hat U(t)\right] \hat U^\dagger(t) \right] \hat U(t) \ket{\psi}
\end{equation}
So under the unitary transformation we have
\begin{equation}
\hat H \rightarrow \hat U(t)\hat H \hat U^\dagger(t) + \mathrm{i} \partial_t\left[\hat U(t)\right] \hat U^\dagger(t).
\end{equation}

Let's now consider the specific transformation given in the main text. We begin with a uniform electric field $\bs{E}$ applied via a scalar potential to give,
\begin{equation}
\hat{\mathcal{H}}' = \hat{\mathcal{H}}_0 - \sum_{\bs{R}, i} q \bs{E}\cdot \bs{R} c^\dagger_{\bs{R}, i}c_{\bs{R}, i}.
\end{equation}
We make a time-dependent unitary transformation of the Hamiltonian. The operator we use is the unitary $\hat{U}(t) =\exp(-\mathrm{i}q\bs{E}\cdot \hat{\bs{R}}t)$. This acts on the creation operator for momentum $\bs{k}$ and orbital $i$ as, $\hat U(t) c^\dagger_{\bs{k}, i} \hat U(t)^\dagger = c^\dagger_{\bs{k}-q\bs{E} t, i}$. Under such a transformation, the Hamiltonian has to transform to the time dependent Hamiltonian $\hat{\mathcal{H}}(t)= \hat U(t) \hat{\mathcal{H}}' \hat U^\dagger(t) + \mathrm{i} \partial_t\left[\hat U(t)\right] \hat U^\dagger(t)$ (see Appendix.~\ref{sec:UnitaryTransformation}). If we explicitly evaluate this last term we obtain,
\begin{align}
\mathrm{i} \partial_t\left[\hat U(t)\right] \hat U^\dagger(t) &= \left(q\bs{E}\cdot\hat{\bs{R}}\right)  \:\hat U(t)\hat U^\dagger(t)\\
&= q\bs{E}\cdot\hat{\bs{R}}\\
&= \sum_{\bs{R}, i}q\bs{E}\cdot {\bs{R}}c^\dagger_{{\bs{R}}, i}c_{{\bs{R}}, i}.
\end{align}
We see that again this term cancels with the scalar potential. Hence the time dependent Hamiltonian is simply $\hat{\mathcal{H}}(t) = \hat U(t) \hat{\mathcal{H}}_0 \hat U^\dagger(t)$.

\newpage

\section{Instantaneous eigenstates of $\hat{\mathcal{H}}(t)$}\label{sec:InstantaneousEigStates}
In this section we verify that the Hamiltonian $\hat{\mathcal{H}}(t)$ is diagonal in the canonical momentum and that the instantaneous eigenstates at fixed canonical momentum $\bs{p}_0$ are
\begin{equation}
\ket{\psi_0(\bs{p}_0, t)} = \hat U(t) \ket{\bs{p}(t) },
\end{equation} 
where $\boldsymbol{p}(t) = \boldsymbol{p}_0 + \sum_\alpha N_\alpha \nu_\alpha q \boldsymbol{E}t$ as discussed in the main text. 

We start by understanding how $\hat U(t)$ transforms $\hat{\mathcal{H}}_0$. Note that it is easiest to work in terms of plane waves (\emph{i.e.} $c^\dagger_{\boldsymbol{k}, i}$ with $i$ an \emph{orbital}) not the Bloch states (\emph{i.e.} $c^\dagger_{\boldsymbol{k}, \alpha}$ with $\alpha$ a \emph{band} label). This is because Bloch states do not transform nicely under the unitary transformation $\hat U(t)$ as,
\begin{align}
\hat Uc^\dagger_{\bs{k}, c}\hat U^\dagger &= \sum_{i} u^{\bs{k}, c}_{i} \hat Uc^\dagger_{\bs{k}, i}\hat U^\dagger\\
&= \sum_{i} u^{\bs{k}, c}_{i} c^\dagger_{\bs{k}-q \bs{E}t, i}\\
&\neq c^\dagger_{\bs{k}-q\bs{E}t, c}.
\end{align}
Therefore we express everything in terms of plane waves which do transform nicely under the unitary $\hat U(t)$. Even the non-interacting ground state we have to write in terms of plane waves since $\hat U(t)$ acts on the ground state when we calculate $\hat{\mathcal{H}}(t)$. Note that we now restrict ourselves to one valence band for notational simplicity however this is not necessary. The ground state can be written as,
\begin{align}
\ket{\mathrm{GS}} &= \prod_{\bs{k}} c^\dagger_{\bs{k}, v} \ket{0}\\
&= \prod_{\bs{k}} \sum_{i} u^{\bs{k}, v}_{i}c^\dagger_{\bs{k},  i} \ket{0}\\
&= \sum_{\bs{n}} \prod_{\bs{k}'}\left( u^{\bs{k}', v}_{\bs{n}[\bs{k}']} \right) \prod_{\bs{k}} c^\dagger_{\bs{k}, \bs{n}[\bs{k}]} \ket{0}\\
&= \sum_{\bs{n}} \psi^{\bs{n}}_{\mathrm{GS}} \ket{\bs{n}} .
\end{align}
We have defined the plane wave basis $\ket{\bs{n}}\equiv\prod_{k} c^\dagger_{\bs{k}, \bs{n}[\bs{k}]} \ket{0} $ with prefactor $\psi^{\bs{n}}_{\mathrm{GS}} \equiv \prod_{\bs{k}'}\left( u^{\bs{k}', v}_{\bs{n}[\bs{k}']} \right)$. The vector $\bs{n}$ has a single entry per momentum $\bs{k}$ that is written as $\bs{n}[\bs{k}]$. This entry specifies the orbital of the plane wave at that momentum. 
Importantly, the $\hat U(t)$ operator acts nicely on the basis states ($\ket{\bs{n}}$) in the thermodynamic limit. In this limit, all the momenta are shifted by some constant amount and you can reorder the product as all continuous momenta exist in the product \emph{i.e.} \begin{align}
 \hat U \ket{\bs{n}} = \prod_{\bs{k}} c^\dagger_{\bs{k}-q \bs{E}t, \bs{n}[\bs{k}]} \ket{0} &=\prod_{\bs{k}} c^\dagger_{\bs{k}, \bs{n}[\bs{k}+q \bs{E}t]} \ket{0}
 \\&\equiv \ket{\hat U\bs{n}}
\end{align} 
where we define the state $\ket{\hat U\bs{n}}$ as the transformed basis state. In the thermodynamic limit all $\hat U$ does is reorder the orbitals associated with each plane wave in $\ket{\bs{n}}$. 

The expression for the bound states we also write in terms of plane waves. In general a plane wave basis for a bound state with $N$ particles takes the form,
\begin{equation}
\ket{(\bs{k}, i), \bs{n}} \equiv c^\dagger_{\bs{k}_1, i_1} c^\dagger_{\bs{k}_2, i_2}\dots c_{\bs{k}_{N-1}, i_{N-1}} c_{\bs{k}_N, i_N} \ket{\bs{n}},\label{eq:basisstateboundstate}
\end{equation}
where the bound state can consist of both particles (creation operators) and holes (annihilation operators). The indices $i_n$ label orbitals (not species). The notation $(\bs{k}, i)$ is shorthand for the explicit list of $\bs{k}$ values and orbitals \emph{i.e.} $(\bs{k}, i) = [(\bs{k}_1, i_1), (\bs{k}_2, i_2), \dots (\bs{k}_N, i_N)]$.

The bound state Hamiltonian is found by projecting the many body Hamiltonian $\hat{\mathcal{H}}_0$ into the basis $\ket{(\bs{k}, i), \bs{n}}$, 
\begin{equation}
\hat{\mathcal{H}}_0 = \sum_{\bs{k},\bs{k}' i, i',  \bs{n},  \bs{n}'} \ket{(\bs{k}, i), \bs{n}}\bra{(\bs{k}', i'), \bs{n}'} \bra{(\bs{k}, i),  \bs{n}}\hat{\mathcal{H}}_0\ket{(\bs{k}', i'),  \bs{n}'}
\end{equation}
Since we have translational symmetry, 
the eigenstates of $\hat{\mathcal{H}}_0$ are labelled by total momentum $\bs{p} = \sum_i \nu_i \bs{k}_i$ where $\nu_i = +1$ for a particle and $\nu_i = -1$ for a hole. The eigenstates of $\hat{\mathcal{H}}_0$ are the $\ket{\bs{p}}$ given in Eq.~\eqref{eq:AppboundstateWavefunction}. We rewrite $\ket{\bs{p}}$ in the plane wave basis as,
\begin{equation}
\ket{\bs{p}} = \mathcal{
N}\sum_{\bs{K} = \bs{p}} \sum_{i, \bs{n}} \Phi^{(\bs{k}, i)} \psi^{\bs{n}}_{\mathrm{GS}} \ket{(\bs{k}, i), \bs{n}}.\label{eq:H0eigenstates}
\end{equation} 
Note that we have only summed over basis states with total momentum $K = \sum_i \nu_i \bs{k}_i$ equal to $\bs{p}$. In addition we have split the wave function into a ground state component ($\psi^{\bs{n}}_{\mathrm{GS}}$) and a bound state component $\Phi^{(\bs{k}, i)}$. We justify this separation using the fact we are in an insulator and so the gap is large compared to the interaction which is binding the particles. We can write an explicit expression for $\Phi^{(\bs{k}, i)}$ using Eq.~\eqref{eq:AppboundstateWavefunction}. It is written in terms of the Bloch functions of the underlying bands as well as $\phi^{(\bs{k})}$,
\begin{align}
\Phi^{(\bs{k}, i)} = \phi^{(\bs{k})} u^{\bs{k}_1, \alpha_1}_{i_1}u^{\bs{k}_2, \alpha_2}_{i_2}\dots \bar u^{\bs{k}_{N-1}, \alpha_{N-1}}_{i_{N-1}} \bar u^{\bs{k}_{N}, \alpha_{N}}_{i_{N}} \label{eq:wavefunctionDecomp}
\end{align}
where $u^{\bs{k_1}, \alpha_1}_{i_1}$ is the eigenstate of the single-electron Bloch Hamiltonian (\emph{i.e.} the cell-periodic function) for momentum $\bs{k_1}$, band/species $\alpha_1$ and orbital $i_1$. The conjugate appears because we can have holes as well as particles in the bound state [\emph{e.g.} both creation and annihilation operators can appear in Eq.~\eqref{eq:basisstateboundstate}].

Next we construct the time dependent Hamiltonian $\hat{\mathcal{H}}(t) = \hat U(t) \hat{\mathcal{H}}_0 \hat U^\dagger(t)$ and work out its instantaneous eigenstates. Since we have written it in terms of plane waves it transforms nicely. First we consider how the unitary acts on the bound state basis,
\begin{equation}
\hat U(t)\ket{(\bs{k}, i), \bs{n}} = \ket{(\bs{k}- q \bs{E} t, i), \hat U\bs{n}}
\end{equation}
where the notation indicates that all $\bs{k}_j \rightarrow \bs{k}_j - q\bs{E}t$. Therefore, 
\begin{align}
\hat{\mathcal{H}}(t) &= \sum_{\bs{k},\bs{k}' i, i',  \bs{n},  \bs{n}'} \hat U(t) \ket{(\bs{k}, i), \bs{n}}\bra{(\bs{k}', i'), \bs{n}'}\hat U^\dagger(t)  \bra{(\bs{k}, i),  \bs{n}}\mathcal{H}_0 \ket{(\bs{k}', i'),  \bs{n}'}\\
&= \sum_{\bs{k},\bs{k}' i, i',  \bs{n},  \bs{n}'} \ket{(\bs{k}-q\bs{E}t, i), \hat U\bs{n}}\bra{(\bs{k}'-q\bs{E}t, i'), \hat U\bs{n}'} \bra{(\bs{k}, i), \bs{n}}\hat{\mathcal{H}}_0 \ket{(\bs{k}', i'),  \bs{n}'}
\\
&= \sum_{\bs{k},\bs{k}' i, i',  \bs{n},  \bs{n}'} \ket{(\bs{k}, i), \bs{n}}\bra{(\bs{k}', i'), \bs{n}'} \bra{(\bs{k}+q\bs{E}t, i),  \hat U^\dagger\bs{n}}\hat{\mathcal{H}}_0 \ket{(\bs{k}'+q\bs{E}t, i'),  \hat U^\dagger\bs{n}'}.
\end{align}
Using the form of the eigenstates of $\hat{\mathcal{H}}_0$ in Eq.~\eqref{eq:H0eigenstates} we can immediately write down the instantaneous eigenstates of $\hat{\mathcal{H}}(t)$ these are, 
\begin{equation}
\ket{\psi_0(\bs{p}_0, t)} = \mathcal{
N}\sum_{\bs{K} = \bs{p}_0}\sum_{i, \bs{n}} \Phi^{(\bs{k}+q\bs{E} t, i)} \psi^{\hat U^\dagger\bs{n}}_{\mathrm{GS}} \ket{(\bs{k}, i), \bs{n}}.
\end{equation}
We note that $\bs{p}_0 = \sum_{i}\nu_i \bs{k}_i$ is conserved and $\bs{p}(t) = \sum_{i}\nu_i (\bs{k}_i+q\bs{E}t)$ is the kinetic momentum which is not conserved. This is exactly the same as the more compact expression in the main text,
\begin{equation}
\ket{\psi_0(\bs{p}_0, t)} = \hat U(t) \ket{\bs{p}(t) }.
\end{equation} 
\newpage
\section{Time derivative of the position operator}\label{sec:PosOperatorDerivative}
We calculate the EOM by finding the time derivative of the expectation value of the position operator $\hat{\bs{R}}^\alpha$. First we must confront a subtlety. For some operator $\hat{\mathcal{O}}$ and state $\ket{\psi}$, if we change the electromagnetic gauge we take $\ket{\psi}\rightarrow \hat U(t) \ket{\psi}$. However we wish expectation values (\emph{i.e.} $\bra{\psi}\hat{\mathcal{O}}\ket{\psi}$) to remain unchanged, this is true for all $\hat{\mathcal{O}}$ if we simultaneously transform the operator $\hat{\mathcal{O}}\rightarrow \hat U(t)\hat{\mathcal{O}}\hat U^\dagger(t)$ when we make an EM gauge transformation. 
The total position operator $\hat{\bs{R}}$ is actually invariant under electromagnetic gauge transformations \emph{i.e.} $\hat{\bs{R}} = \hat U(t)\hat{\bs{R}}\hat U^\dagger(t)$ so we don't need to bother to do this for the single-electron case. However, this is not true for the position operators for species $\alpha$ ($\hat{\bs{R}}^\alpha$). Since we have changed the electromagnetic gauge using $\hat U(t)$, the species position operator transforms to $\hat{U}(t)\hat{\bs{R}}^\alpha \hat U^\dagger(t)$.

Next we show that the (species) position operator acts like a derivative on the bound state wave function. Consider the instantaneous eigenstate,
\begin{equation}
\ket{\psi_0(\bs{p}_0, t)} = \hat U(t) \ket{\bs{p}(t)}
\end{equation}
and $\ket{\psi(\bs{p}_0, t)} = \gamma({\bs{p}_0, t})\ket{\psi_0(\bs{p}_0, t)}$.
We now show that,
\begin{align}
\frac{\mathrm{d}}{\mathrm{d}t}\left[\bra{\psi(\bs{p}_0, t)}\hat U(t)\hat{\bs{R}}^{\alpha}\hat U^\dagger(t)\ket{\psi(\bs{p}_0, t)}\right] &=\mathrm{i}\frac{\mathrm{d}}{\mathrm{d}t}\left(\bar\gamma \nabla_{\bs{p}_0}\gamma\right)+\frac{\mathrm{d}}{\mathrm{d}t}\bs{\mathcal{A}}^\alpha[\bs{p}(t)].\label{eq:proveDerivative}
\end{align}
We first simplify the LHS of Eq.~\eqref{eq:proveDerivative}. We use,
\begin{align}
\hat U^\dagger(t) \ket{\psi_0(\bs{p}_0, t)} 
&= \ket{\bs{p}(t)},
\end{align}
In addition to simplify the notation we define $\bs{k}_{n}(t) \equiv \bs{k}_{n} + q\bs{E} t$.

A direct way to show Eq.~\eqref{eq:proveDerivative} is to first use the periodic position operators for each species. This simplifies the calculation since the normal position operator $\bs{R}^{\alpha}$ is only well defined in the thermodynamic limit. We write the periodic position operator as $\hat Z^{\alpha}_i$ for particles of species $\alpha$ for direction $ i\in \{x, y, z\}$. Without loss of generality we restrict to the $x$ direction. If species $\alpha$ are particles (not holes) then we use the number operator projected to the $\alpha$ species, $\hat n_{\bs{R}}^\alpha = \sum_{\bs{k}, \bs{k}', i} \bar\psi^{\bs{k}, \alpha}_{\bs{R}, i} \psi^{\bs{k}', \alpha}_{\bs{R}, i} c^\dagger_{\bs{k}, \alpha} c_{\bs{k}', \alpha}$. The periodic position operator for the average position of particles of species $\alpha$ in the $x$ direction is,
\begin{align}
\hat Z^{\alpha}_{x} &= \frac{1}{N_\alpha}\sum_{\bs{R}} e^{\mathrm{i}\bs{\Delta}_{x}\cdot\bs{R}} \hat n_{\bs{R}}^\alpha\\
&=  \frac{1}{N_\alpha}\frac{1}{\mathcal{V}}\sum_{\bs{k}, \bs{k}', \bs{R}, i} e^{\mathrm{i}(\bs{\Delta}_{x} -\bs{k}+\bs{k}')\cdot\bs{R}}\:\bar u^{\bs{k}, \alpha}_{i} u^{\bs{k}', \alpha}_{i} c^\dagger_{\bs{k}, \alpha} c_{\bs{k}', \alpha}
\\
&=  \frac{1}{N_\alpha}\sum_{\bs{k}, i}\:\bar u^{\bs{k}+\bs{\Delta}_{x}, \alpha}_{i} u^{\bs{k}, \alpha}_{i} c^\dagger_{\bs{k}+\bs{\Delta}_{x}, \alpha} c_{\bs{k}, \alpha}\\
&=  \frac{1}{N_\alpha}\sum_{\bs{k}}\:\braket{u^{\bs{k}+\bs{\Delta}_{x}, \alpha}| u^{\bs{k}, \alpha} }c^\dagger_{\bs{k}+\bs{\Delta}_{x}, \alpha} c_{\bs{k}, \alpha}
\end{align}
where $\bs{\Delta}_{x} = \frac{2\pi}{L} \hat{\bs{x}}$ for system size $L$.

We want to take the thermodynamic limit $L\rightarrow\infty$ in order to get the position operators as derivatives. By performing a Taylor series for large $L$ (and therefore small $|\bs{\Delta}_x|$) we can show that,
\begin{equation}
 -\mathrm{i}\lim_{L\rightarrow\infty}\left(\frac{\hat Z^{\alpha}_{x}-1}{|\bs{\Delta}_x|}\right) = \hat R_x^{\alpha},\label{eq:limitofZ}
\end{equation}
We have defined $\hat R_x^{\alpha}$ as the position operator for the $x$-direction for species $\alpha$, \emph{i.e.} taking only the $x$ component position of $\hat{\bs{R}}^\alpha$ from Eq.~\eqref{eq:particlePositionOperator}. The expectation value of $\hat{Z}^{\alpha}_x$ for a single bound state wave function [$\bra{\bs{p}(t)}\hat{Z}^{\alpha}_x\ket{\bs{p}(t)}$] vanishes because $Z^{\alpha}_x$ shifts the total momentum of the state. Hence we have to consider a wave-packet with weighting coefficients $a_{\bs{p}}$,
\begin{align}
\ket{\psi} &= \int_{\mathrm{BZ}} {\mathrm{d}^d\bs{p}_0} \:
a_{\bs{p}_0}\hat U^\dagger(t)\ket{\psi_0(\bs{p}_0, t)}\\
&= \int_{\mathrm{BZ}} {\mathrm{d}^d\bs{p}_0} \:
a_{\bs{p}_0}\gamma(\bs{p}_0, t)\ket{\bs{p}(t)},
\end{align}
with $ \int_{\mathrm{BZ}} {\mathrm{d}^d\bs{p}_0} \:
|a_{\bs{p}_0}|^2 = 1$.
To calculate $\bra{\psi}\hat Z^{\alpha}_x\ket{\psi}$ we use,
\begin{equation}
\hat Z^{\alpha}_x c^\dagger_{\bs{k}, \beta} = c^\dagger_{\bs{k}, \beta} \hat Z^{\alpha}_x + \delta_{\alpha, \beta} \frac{1}{N_\alpha}\braket{u^{\bs{k}+\bs{\Delta}_{x}, \alpha}| u^{\bs{k}, \alpha}}  c^\dagger_{\bs{k}+\bs{\Delta}_x, \alpha}
\end{equation}
Therefore,
\begin{align}
\hat Z^{\alpha}_x \ket{\bs{p}(t)} &= \frac{1}{N_\alpha}\mathcal{
N}\sum _{n: \alpha_n = \alpha}\bigg[\sum_{\bs{K} = \bs{p}_0} \gamma(\bs{p}_0, t)\phi^{(\bs{k}(t))}\braket{u^{\bs{k}_n(t)+\bs{\Delta}_{x}, \alpha}| u^{\bs{k}_n(t), \alpha}}\nonumber \\ &\quad\quad\quad\quad\quad\quad \cdot c^\dagger_{\bs{k}_1(t), \alpha_1}\dots c^\dagger_{\bs{k}_n(t)+\bs{\Delta}_x, \alpha_n}\dots c_{\bs{k}_{N}(t), \alpha_{N}}\ket{\mathrm{GS}}\bigg].
\end{align}
The summation $\sum _{n:\alpha_n = \alpha}$ is over all particles/holes (labelled by $n$) within the bound state which are of species $\alpha$. The state ($\hat Z^{\alpha}_x \hat U^\dagger \ket{\psi(\bs{p}_0, t)}$) only has non-zero overlap with the state $\ket{\psi(\bs{p}_0+\bs{\Delta}_x, t)}$. This means that when we calculate the expectation value $\bra{\psi} \hat Z^{\alpha}_x \ket{\psi}$, the only contributing terms are 
\begin{align}
\bra{\psi(\bs{p}_0+\bs{\Delta}_x, t)}\hat U\hat Z^{\alpha}_x \hat U^\dagger \ket{\psi(\bs{p}_0, t)} &= \frac{1}{N_\alpha}\sum _{n:\alpha_n = \alpha}\bigg[\sum_{\bs{K} = \bs{p}_0} \bar \gamma(\bs{p}_0+\bs{\Delta}_x, t)\gamma(\bs{p}_0, t)\\ &\nonumber\quad\quad\quad\quad\cdot\bar \phi^{(\dots, \bs{k}_n(t)+\bs{\Delta}_x, \dots)}\phi^{(\dots, \bs{k}_n(t), \dots)}\braket{u^{\bs{k}_n(t)+\bs{\Delta}_{x}, \alpha}| u^{\bs{k}_n(t), \alpha}}\bigg].
\end{align}
Note that we have used Wick's theorem and the anti-symmetry of the bound state wave function to simplify this. See Sec.~\ref{sec:BoundStateBerry} for details. 

Therefore,
\begin{align}
\bra{\psi} \hat Z^{\alpha}_x \ket{\psi} &= \sum_{\bs{p}_0} \bar{a}_{\bs{p_0}+\bs{\Delta}_x} a_{\bs{p_0}}\bra{\psi_0(\bs{p}_0+\bs{\Delta}_x, t)}\hat U\hat Z^{\alpha}_x \hat U^\dagger \ket{\psi_0(\bs{p}_0, t)}\\
&= \sum_{\bs{p}_0} \bar{a}_{\bs{p_0}+\bs{\Delta}_x} a_{\bs{p_0}}\frac{1}{N_\alpha}\sum _{n:\alpha_n = \alpha}\bigg[\sum_{\bs{K} = \bs{p}_0} \bar\gamma(\bs{p}_0+\bs{\Delta}_x, t)\gamma(\bs{p}_0, t) \\&\nonumber\quad\quad\quad\bar \phi^{(\dots, \bs{k}_n(t)+\bs{\Delta}_x, \dots)}\phi^{(\dots, \bs{k}_n(t), \dots)}\braket{u^{\bs{k}_n(t)+\bs{\Delta}_{x}, \alpha}| u^{\bs{k}_n(t), \alpha}}\bigg].
\end{align}
We first simplify this by identifying that each term in the sum over $n: \alpha_n = \alpha$ is identical so we can simply write,
\begin{align}
\bra{\psi} \hat Z^{\alpha}_x \ket{\psi} 
&= \sum_{\bs{p}_0} \bar{a}_{\bs{p_0}+\bs{\Delta}_x} a_{\bs{p_0}}\bigg[\sum_{\bs{K} = \bs{p}_0} \bar\gamma(\bs{p}_0+\bs{\Delta}_x, t)\gamma(\bs{p}_0, t) \\&\nonumber\quad\quad\quad\bar \phi^{(\dots, \bs{k}_n(t)+\bs{\Delta}_x, \dots)}\phi^{(\dots, \bs{k}_n(t), \dots)}\braket{u^{\bs{k}_n(t)+\bs{\Delta}_{x}, \alpha}| u^{\bs{k}_n(t), \alpha}}\bigg].
\end{align}
for some particle labelled by $n$ in the bound state which is of species $\alpha$ \emph{i.e.} $\alpha_n = \alpha$. The expectation value of $\hat{R}^\alpha_x$ is calculated by taking the thermodynamic limit as in Eq.~\eqref{eq:limitofZ}. We use $\phi^{(\dots, \bs{k}_n(t)+\bs{\Delta}_x, \dots)}\approx \phi^{(\dots, \bs{k}_n(t), \dots)} +\bs{\Delta}_x\cdot \nabla_{\bs{k}_n}\phi^{(\dots, \bs{k}_n(t), \dots)}$ and similar Taylor expansions for $\ket{u^{\bs{k}_n(t)+\bs{\Delta}_x, \alpha}}$ and $\gamma(\bs{p}_0+\bs{\Delta}_x, t)$. Plugging this in we get,
\begin{align}
\bra{\psi} \hat Z^{\alpha}_x \ket{\psi} 
\approx &\sum_{\bs{p}_0} \left(\bar{a}_{\bs{p_0}}+ \bs{\Delta}_x\cdot \nabla_{\bs{p}_0}\bar{a}_{\bs{p_0}}\right)a_{\bs{p_0}}\\&\nonumber\cdot\bigg[\sum_{\bs{K} = \bs{p}_0} \bigg(\bar\gamma(\bs{p}_0, t)+ \bs{\Delta}_x\cdot \nabla_{\bs{p}_0}\bar\gamma(\bs{p}_0, t) \bigg)\gamma(\bs{p}_0, t) \\&\nonumber\quad \quad\cdot\bigg(\bar \phi^{(\dots, \bs{k}_n(t), \dots)}+ \bs{\Delta}_x\cdot \nabla_{\bs{k}_n}\bar \phi^{(\dots, \bs{k}_n(t), \dots)}\bigg)\phi^{(\dots, \bs{k}_n(t), \dots)}\\ &\nonumber\quad\quad\cdot\left(\bra{u^{\bs{k}_n(t), \alpha}}+ \bs{\Delta}_x\cdot \nabla_{\bs{k}_n}\bra{u^{\bs{k}_n(t), \alpha}}\right)\ket{u^{\bs{k}_n(t), \alpha}}\bigg].
\end{align}
Since we are interested in the time derivative of the position, we can throw  away terms which are constant. To first order in $\bs{\Delta}_x$ we can therefore get rid of the $\nabla_{\bs{p}_0}\bar{a}_{\bs{p_0}}$ term. Using this we find,
\begin{align}
\frac{\mathrm{d}}{\mathrm{d}t}\bra{\psi} \hat Z^{\alpha}_x \ket{\psi} 
=&\frac{\mathrm{d}}{\mathrm{d}t}\bigg[\sum_{\bs{p}_0} |a_{\bs{p}_0}|^2\sum_{\bs{K} = \bs{p}_0} \bigg(\bar\gamma(\bs{p}_0, t)+ \bs{\Delta}_x\cdot \nabla_{\bs{p}_0}\bar\gamma(\bs{p}_0, t) \bigg)\gamma(\bs{p}_0, t)\\&\quad\quad\cdot\left(|\phi^{(\dots, \bs{k}_n(t), \dots)}|^2+ \bs{\Delta}_x\cdot \nabla_{\bs{k}_n}\left(\bar \phi^{(\dots, \bs{k}_n(t), \dots)}\right)\phi^{(\dots, \bs{k}_n(t), \dots)}\right)\\ &\nonumber\quad\quad\cdot\left(1+ \bs{\Delta}_x\cdot \nabla_{\bs{k}_n}\left(\bra{u^{\bs{k}_n(t), \alpha}}\right)\ket{u^{\bs{k}_n(t), \alpha}}\right)+\mathcal{O}(|\bs{\Delta}_x|^2)\bigg]\\
&=\frac{\mathrm{d}}{\mathrm{d}t}\bigg\{\sum_{\bs{p}_0} |a_{\bs{p}_0}|^2\bigg[1+\bs{\Delta}_x\cdot \nabla_{\bs{p}_0}[\bar\gamma(\bs{p}_0, t) ]\gamma(\bs{p}_0, t) \\&\nonumber\quad\quad\quad\quad\quad\quad\quad\quad+\sum_{\bs{K} = \bs{p}_0} \bigg( \bs{\Delta}_x\cdot \nabla_{\bs{k}_n}\left(\bar \phi^{(\dots, \bs{k}_n(t), \dots)}\right)\phi^{(\dots, \bs{k}_n(t), \dots)}\\ &\nonumber\quad\quad\quad\quad\quad\quad\quad\quad\quad\quad+ |\phi^{(\dots, \bs{k}_n(t), \dots)}|^2\bs{\Delta}_x\cdot \nabla_{\bs{k}_n}\left(\bra{u^{\bs{k}_n(t), \alpha}}\right)\ket{u^{\bs{k}_n(t), \alpha}}\bigg)+\mathcal{O}(|\bs{\Delta}_x|^2)\bigg]\bigg\}\\
&=\frac{\mathrm{d}}{\mathrm{d}t}\bigg\{1 - \sum_{\bs{p}_0} |a_{\bs{p}_0}|^2\bigg\{\bar\gamma(\bs{p}_0, t) \bs{\Delta}_x\cdot \nabla_{\bs{p}_0}\gamma(\bs{p}_0, t) \\&\nonumber\quad\quad\quad\quad\quad\quad\quad\quad\quad\quad+ \sum_{\bs{K} = \bs{p}_0} \bigg(\bar \phi^{(\dots, \bs{k}_n(t), \dots)}\bs{\Delta}_x\cdot \nabla_{\bs{k}_n}\phi^{(\dots, \bs{k}_n(t), \dots)}\\ &\nonumber\quad\quad\quad\quad\quad\quad\quad\quad\quad\quad+ |\phi^{(\dots, \bs{k}_n(t), \dots)}|^2\bra{u^{\bs{k}_n(t), \alpha}}\bs{\Delta}_x\cdot \nabla_{\bs{k}_n}\ket{u^{\bs{k}_n(t), \alpha}}\bigg)\bigg\}+\mathcal{O}(|\bs{\Delta}_x|^2)\bigg\}.
\end{align}
We now assume we can consider only one momentum in the wave packet so we restrict to one value of $\bs{p}_0$. We could equally continue with the full wave packet, we would get the sum over the equations of motion of each momentum $\bs{p}_0$ weighted by $|a_{\bs{p}_0}|^2$. To further simplify the notation we now write $\bs{\Delta}_x\cdot \nabla_{\bs{k}} = |\bs{\Delta}_x|\partial_{\bs{k}^x}$ where $\bs{k}^x$ is the $x^{\mathrm{th}}$ component of momentum $\bs{k}$. We use Eq.~\eqref{eq:limitofZ} to get $\bra{\psi} \hat R^{\alpha}_x \ket{\psi} $,
\begin{align}
\frac{\mathrm{d}}{\mathrm{d}t}\bra{\psi} \hat R^{\alpha}_x \ket{\psi} &= \mathrm{i}\frac{\mathrm{d}}{\mathrm{d}t} \left(\bar\gamma \partial_{\bs{p}_0^x}\gamma \right)+\frac{\mathrm{d}}{\mathrm{d}t}\bigg[\sum_{\bs{K} = \bs{p}_0} \bigg(\bar \phi^{(\dots, \bs{k}_n(t), \dots)}\mathrm{i}\partial_{\bs{k}_n^x}\phi^{(\dots, \bs{k}_n(t), \dots)}\\ &\nonumber\quad\quad\quad\quad\quad\quad\quad\quad+ |\phi^{(\dots, \bs{k}_n(t), \dots)}|^2\bra{u^{\bs{k}_n(t), \alpha}}\mathrm{i}\partial_{\bs{k}_n^x}\ket{u^{\bs{k}_n(t), \alpha}}\bigg)\bigg].
\end{align}
Since this equation holds regardless of what direction position operator we had used (\emph{e.g.} $x, y, z$), we can now return to the vector position operator $\hat{\bs{R}}^{\alpha}$ [defined in Eq.~\eqref{eq:particlePositionOperator}].
\begin{align}
\frac{\mathrm{d}}{\mathrm{d}t}\bra{\psi} \hat{\bs{R}}^{\alpha} \ket{\psi} &= \mathrm{i}\frac{\mathrm{d}}{\mathrm{d}t} \left(\bar\gamma \nabla_{\bs{p}_0}\gamma \right)+\frac{\mathrm{d}}{\mathrm{d}t}\bigg[\sum_{\bs{K} = \bs{p}_0} \bigg(\bar \phi^{(\dots, \bs{k}_n(t), \dots)}\mathrm{i}\nabla_{\bs{k}_n}\phi^{(\dots, \bs{k}_n(t), \dots)}\\ &\nonumber\quad\quad\quad\quad\quad\quad\quad\quad+ |\phi^{(\dots, \bs{k}_n(t), \dots)}|^2\bra{u^{\bs{k}_n(t), \alpha}}\mathrm{i}\nabla_{\bs{k}_n}\ket{u^{\bs{k}_n(t), \alpha}}\bigg)\bigg].
\end{align}
We note that the second term in this expression is exactly our expression for $\bs{\mathcal{A}}^\alpha(\bs{p})$ evaluated at $\bs{p}(t)$ so that,
\begin{align}
\frac{\mathrm{d}}{\mathrm{d}t}\bra{\psi} \hat{\bs{R}}^{\alpha} \ket{\psi} &= \mathrm{i}\frac{\mathrm{d}}{\mathrm{d}t} \left(\bar\gamma \nabla_{\bs{p}_0}\gamma \right)+\frac{\mathrm{d}}{\mathrm{d}t}\bs{\mathcal{A}}^\alpha[\bs{p}(t)].
\end{align}
We therefore have proved the expression stated in the main text [Eq.~\eqref{eq:proveDerivative}].

Lastly, the above derivation assumed species $\alpha$ was particle not hole-like. The periodic position operator for holes is,
\begin{align}
\hat Z^{\alpha}_{x} &= \frac{1}{N_\alpha}\sum_{\bs{R}} e^{\mathrm{i}\bs{\Delta}_{x}\cdot\bs{R}} \hat n_{\bs{R}}^\alpha\\
&=  \frac{1}{N_\alpha}\frac{1}{\mathcal{V}}\sum_{\bs{k}, \bs{k}', \bs{R}, i} e^{\mathrm{i}(\bs{\Delta}_{x} +\bs{k}-\bs{k}')\cdot\bs{R}}\: u^{\bs{k}, \alpha}_{i} \bar u^{\bs{k}', \alpha}_{i} c_{\bs{k}, \alpha} c^\dagger_{\bs{k}', \alpha}
\\
&=  \frac{1}{N_\alpha}\sum_{\bs{k}, i}  u^{\bs{k}-\bs{\Delta}_{x}, \alpha}_{i} \bar u^{\bs{k}, \alpha}_{i} c_{\bs{k}-\bs{\Delta}_{x}, \alpha} c^\dagger_{\bs{k}, \alpha}
\\
&=  \frac{1}{N_\alpha}\sum_{\bs{k}}\:\braket{u^{\bs{k}, \alpha}| u^{\bs{k}-\bs{\Delta}_{x}, \alpha} }c_{\bs{k}-\bs{\Delta}_{x}, \alpha} c^\dagger_{\bs{k}, \alpha}
\end{align}
We see that for a hole the periodic position operator gives a negative momentum shift this means that expression for the time derivative is unchanged,
\begin{align}
\frac{\mathrm{d}}{\mathrm{d}t}\bra{\psi} \hat{\bs{R}}^{\alpha} \ket{\psi} &= \mathrm{i}\frac{\mathrm{d}}{\mathrm{d}t} \left(\bar\gamma \nabla_{\bs{p}_0}\gamma \right)+\frac{\mathrm{d}}{\mathrm{d}t}\bs{\mathcal{A}}^\alpha[\bs{p}(t)],
\end{align}
but now using the hole Berry connection $\bs{\mathcal{A}}^\alpha[\bs{p}(t)]$.

Using the periodic position operator allows us to see that our EOM are exact in the thermodynamic limit but have corrections of order $\Delta \sim 1/L$ for finite systems.
\newpage
\section{Proof that ground state time dependence does not matter}\label{sec:groundstatedependencedropsout}
We show that,
\begin{equation}
\mathrm{i}\nabla_{\bs{p}}\left(\braket{\psi_0| \partial_t\psi_0}\right) =  \nabla_{\bs{p}}\left[\sum_\beta N_\beta \nu_\beta q \bs{E}\cdot \bs{\mathcal{A}}^\beta(\bs{p})  \right],
\end{equation}
First we use the form of the instantaneous eigenstates under time evolution,
\begin{equation}
\ket{\psi_0(\bs{p}_0, t)} = \mathcal{N}\sum_{\bs{K} = \bs{p}_0}\sum_{i, \bs{n}} \Phi^{(\bs{k}+q\bs{E} t, i)} \psi^{\hat U^\dagger\bs{n}}_{\mathrm{GS}} \ket{(\bs{k}, i), \bs{n}}.
\end{equation}
We calculate the time derivative as,
\begin{align}
\mathrm{i}\braket{\psi_0| \partial_t\psi_0} &= \sum_{\bs{K} = \bs{p}_0}\sum_{ i, \bs{n}} \bar\Phi^{(\bs{k}+q\bs{E} t,  i)} \bar\psi^{\hat U^\dagger\bs{n}}_{\mathrm{GS}}  \mathrm{i}\frac{d}{dt}\left[\Phi^{(\bs{k}+q\bs{E} t,  i)} \psi^{\hat U^\dagger\bs{n}}_{\mathrm{GS}} \right]\\
&= \sum_{\bs{K} = \bs{p}_0}\sum_{ i} \bar\Phi^{(\bs{k}+q\bs{E} t,  i)}  \mathrm{i}\frac{d}{dt}\left[\Phi^{(\bs{k}+q\bs{E} t, i)}  \right] +\sum_{\bs{n}}\psi^{\hat U^\dagger\bs{n}}_{\mathrm{GS}}\mathrm{i}\frac{d}{dt}\psi^{\hat U^\dagger\bs{n}}_{\mathrm{GS}}\\
&= \sum_{\bs{K} = \bs{p}_0}\sum_{i, \beta} \bar\Phi^{(\bs{k}+q\bs{E} t, i)}   q\bs{E}\cdot \sum_{n: \alpha_n = \beta} \mathrm{i}\nabla_{\bs{k}_n}\left[\Phi^{(\bs{k}+q\bs{E} t, i)} \right] +\sum_{\bs{n}}\psi^{\hat U^\dagger\bs{n}}_{\mathrm{GS}}\mathrm{i}\frac{d}{dt}\psi^{\hat U^\dagger\bs{n}}_{\mathrm{GS}}\\
&= \sum_\beta N_\beta \nu_\beta q \bs{E}\cdot \bs{\mathcal{A}}^\beta(\bs{p})  +\sum_{\bs{n}}\psi^{\hat U^\dagger\bs{n}}_{\mathrm{GS}}\mathrm{i}\frac{d}{dt}\psi^{\hat U^\dagger\bs{n}}_{\mathrm{GS}}\label{eq:groundstateTimedependence}.
\end{align}

We therefore obtain,
\begin{equation}
\mathrm{i}\nabla_{\bs{p}}\left(\braket{\psi_0| \partial_t\psi_0}\right) =  \nabla_{\bs{p}}\left[\sum_\beta N_\beta \nu_\beta q \bs{E}\cdot \bs{\mathcal{A}}^\beta(\bs{p})  \right],
\end{equation}
since the ground state term in Eq.~\eqref{eq:groundstateTimedependence} is independent of the momentum of the bound state $\bs{p}$ it vanishes under the momentum derivative $\nabla_{\bs{p}}$. 

\newpage
\section{Difference between equations of motion of different species}\label{sec:BoundStatesStayBound}
The equation of motion for species $\alpha$ is,
\begin{equation}
\frac{\mathrm{d}\langle \hat{\bs{R}}^\alpha\rangle}{\mathrm{d}t}  =  \nabla_{\bs{p}}E - \dot{\bs{p}}\times \bs{\Omega}^{\alpha}(\bs{p}) +  \nabla_{\bs{p}}\bigg[q \bs{E}\cdot\sum_\beta N_\beta\nu_\beta \bs{\mathcal{D}}^{\alpha, \beta}(\bs{p})\bigg].
\end{equation}
We calculate the time dependence of the difference in positions between two species,
\begin{align}
\frac{\mathrm{d}\langle \hat{\bs{R}}^\alpha\rangle}{\mathrm{d}t} -\frac{\mathrm{d}\langle \hat{\bs{R}}^\beta\rangle}{\mathrm{d}t}  &=  \dot{\bs{p}}\times [\bs{\Omega}^{\beta}(\bs{p})- \bs{\Omega}^{\alpha}(\bs{p})] +  \nabla_{\bs{p}}\bigg[q \bs{E}\cdot\sum_\gamma N_\gamma\nu_\gamma\left(\bs{\mathcal{D}}^{\alpha, \gamma}(\bs{p}) -\bs{\mathcal{D}}^{\beta, \gamma}(\bs{p})\right)\bigg] \\
&=  \dot{\bs{p}}\times [\bs{\Omega}^{\beta}(\bs{p})- \bs{\Omega}^{\alpha}(\bs{p})] +  \nabla_{\bs{p}}\bigg[\sum_\gamma N_\gamma\nu_\gamma q \bs{E}\cdot\bs{\mathcal{D}}^{\alpha, \beta}(\bs{p})\bigg] \\
&=  \dot{\bs{p}}\times [\bs{\Omega}^{\beta}(\bs{p})- \bs{\Omega}^{\alpha}(\bs{p})] +  \nabla_{\bs{p}}\bigg[\dot{\bs{p}}\cdot\bs{\mathcal{D}}^{\alpha, \beta}(\bs{p})\bigg] \\
&=  -\dot{\bs{p}}\times [\nabla_{\boldsymbol{p}}\times \bs{\mathcal{D}}^{\alpha, \beta}(\bs{p})] +  \nabla_{\bs{p}}\bigg[\dot{\bs{p}}\cdot\bs{\mathcal{D}}^{\alpha, \beta}(\bs{p})\bigg]
\\
&=  \left(\dot{\bs{p}}\cdot \nabla_{\boldsymbol{p}} \right)\bs{\mathcal{D}}^{\alpha, \beta}(\bs{p})
\\
&=  \frac{\mathrm{d}}{\mathrm{d}t}\bs{\mathcal{D}}^{\alpha, \beta}(\bs{p}).
\end{align}
Therefore integrating the EOM gives
\begin{equation}
\langle \hat{\bs{R}}^\alpha\rangle(t) - \langle \hat{\bs{R}}^\beta\rangle(t) = \bs{\mathcal{D}}^{\alpha, \beta}[\bs{p}(t)],
\end{equation}
recalling that $\boldsymbol{p}(t) = \boldsymbol{p}_0 + \sum_\alpha N_\alpha \nu_\alpha q \boldsymbol{E}t$. The bound state therefore remains bound when evolved according to the EOM. This is because the $\bs{\mathcal{D}}^{\alpha, \beta}[\bs{p}(t)]$ is (by definition) finite everywhere in the BZ for a bound state.

\newpage
\section{Trions in magic-angle twisted-bilayer graphene}\label{sec:TrionsMATBG}
\begin{figure}[h]
    \centering
    \includegraphics[width=0.9\linewidth]{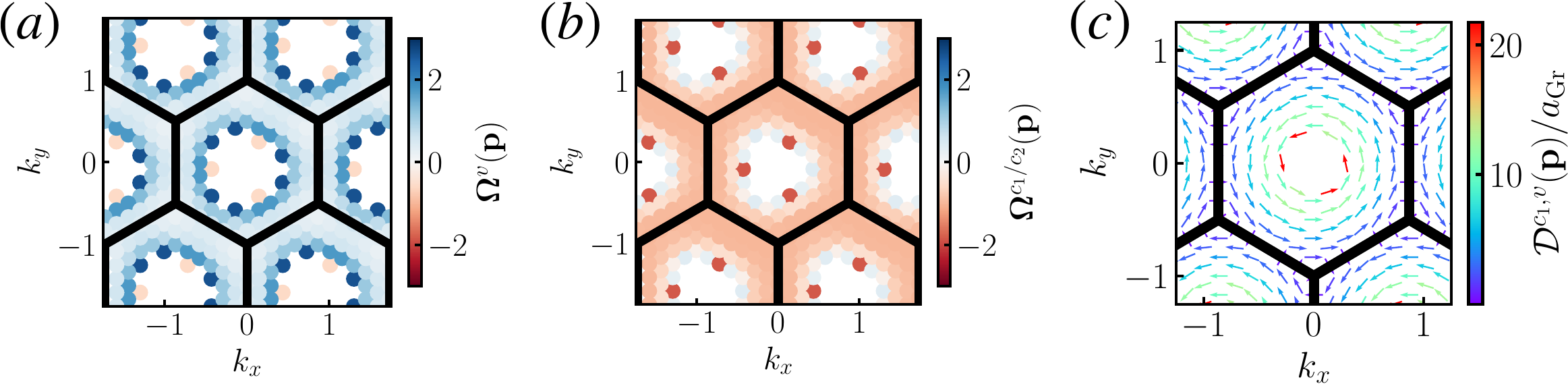}
    \caption{Berry curvartures and QGD for trions in MATBG. The $\Gamma$ point is not gapped from the continuum so we do not plot these regions. $(a)$ Berry curvature for holes $\bs{\Omega}^{v}(\bs{p})$ $(b)$ Berry curvature for the electrons $\bs{\Omega}^{c_1/c_2}(\bs{p})$. The two possible electron Berry curvatures are equal in the chiral limit. $(c)$ The QGD $\bs{\mathcal{D}}^{c_1/c_2, v}(\bs{p})$, again these two QGDs are equal in the chiral limit.}
    \label{fig:TrionsTBGBerryCurvature}
\end{figure}
We study trions in the chiral limit of magic-angle twisted-bilayer graphene (MATBG)~\cite{TBGChiral, TBGII, TBGIII} following the approach in Ref.~\cite{trionsSchindler} at a gate distance of $\xi = 20\:\mathrm{nm}$. We calculate the Berry curvature using the gauge invariant formulation in terms of Eq.~\eqref{eq:WilsonLoop}. This is plotted in Fig.~\ref{fig:TrionsTBGBerryCurvature}. We see that the hole Berry curvature is much smaller compared to the electron Berry curvatures away from the $\Gamma$ point. We also calculate the two independent dipoles which are found to be equal in the simplified chiral limit see Fig.~\ref{fig:TrionsTBGBerryCurvature}.

The equations of motion for trions are,
\begin{align}
\frac{\mathrm{d}\langle \hat{\bs{R}}^{c_1/c_2}\rangle}{\mathrm{d}t} &= e\bs{E}\times \bs{\Omega}^{c_1/c_2}(\bs{p}) + \nabla_{\bs{p}}\left[e\bs{E}\cdot\bs{\mathcal{D}}^{c_1/c_2, v}(\bs{p})\right]\\
\frac{\mathrm{d}\langle \hat{\bs{R}}^{v}\rangle}{\mathrm{d}t} &= e\bs{E}\times \bs{\Omega}^{v}(\bs{p}) + \nabla_{\bs{p}}\left[2e\bs{E}\cdot\bs{\mathcal{D}}^{c_1/c_2, v}(\bs{p})\right].
\end{align}
The resulting dynamics are shown in the main text Fig. 1. Note we assume the dispersion effects can be neglected. As shown in Fig. 1(a) in the main text the trion band is very close to flat at the edge of the BZ. 
\end{document}